\documentclass[aps,prx,twocolumn,superscriptaddress,floatfix,nofootinbib,prx]{revtex4}

\usepackage[usenames,dvipsnames]{color}
\usepackage[colorlinks=true,citecolor=blue,linkcolor=magenta]{hyperref}

\usepackage{graphicx,amsfonts,amssymb,amsmath,hyperref,hypcap,enumerate,epsfig}
\usepackage{xcolor}
\usepackage[caption=false]{subfig}

\newcommand{\bit}{\begin{itemize}}
\newcommand{\eit}{\end{itemize}}

\newcommand{\f}{\frac}
\renewcommand{\>}{\right\rangle}
\newcommand{\<}{\left\langle}
\newcommand{\ba}{\begin{align}}
\newcommand{\ea}{\end{align}}
\newcommand{\be}{\begin{equation}}
\newcommand{\ee}{\end{equation}}
\newcommand{\bi}{\begin{itemize}}
\newcommand{\ei}{\end{itemize}}
\newcommand{\lf}{\left(}
\newcommand{\ri}{\right)}
\newcommand{\dd}{\mathrm{d}}

\newcommand{\tr}{\operatorname{tr}}

\newcommand{\blue}{}

\newcommand{\bra}[1]{\< #1 \right|}
\newcommand{\ket}[1]{\left| #1 \>}

\newcommand{\opket}[1]{|\hspace{-0.15mm}| #1 \rangle\hspace{-0.5mm}\rangle}
\newcommand{\opbra}[1]{\langle\hspace{-0.5mm}\langle #1 |\hspace{-0.15mm}|}

\newcommand{\sspr}{s_\text{spread}}
\newcommand{\seq}{s_\text{eq}}
\newcommand{\lt}{\mathcal{E}}

\begin{document}

\author{Cheryne Jonay}
\affiliation{Department of Physics, Princeton University, Princeton, NJ 08544, USA}
\affiliation{Perimeter Institute for Theoretical Physics, 31 Caroline Street North, Waterloo, ON N2L 2Y5, Canada}
\author{David A. Huse}
\affiliation{Department of Physics, Princeton University, Princeton, NJ 08544, USA}
\author{Adam Nahum}
\affiliation{
Theoretical Physics, Oxford University, 1 Keble Road, Oxford OX1 3NP, United Kingdom}

\title{
Coarse-grained dynamics of operator and state entanglement
}
\begin{abstract}
We give a detailed theory for the leading coarse-grained dynamics of entanglement entropy of states and of operators in generic  short-range interacting quantum many-body systems.
This includes operators spreading under Heisenberg time evolution, which we find are much less entangled than ``typical'' operators of the same spatial support. 
Extending previous conjectures based on random circuit dynamics, we provide evidence that the leading-order entanglement dynamics of a given chaotic system are  determined by a  function $\lt(\vec v)$, which is model-dependent, but which we argue  satisfies certain general constraints.
 In a minimal membrane picture,  $\lt(\vec v)$ is the ``surface tension'' of the membrane and is a function of the membrane's orientation $\vec v$ in spacetime.  For one-dimensional (1D) systems this surface tension is related by a Legendre transformation to an entanglement entropy growth rate  $\Gamma(\partial S/\partial x)$ which depends on the spatial ``gradient'' of the entanglement entropy $S(x,t)$ across the cut at position $x$.
We show how to extract the entanglement growth functions numerically in 1D at infinite temperature using the concept of the operator entanglement of the time evolution operator, and we discuss possible universality of $\lt$ at low temperatures.
Our theoretical ideas are tested against and informed by numerical results for a quantum-chaotic 1D spin Hamiltonian.
These results are relevant to the broad class of chaotic many-particle systems or field theories with spatially local interactions, both in 1D and above.

\end{abstract}

\maketitle

\section{Introduction}

The dynamics of entanglement and operator spreading in quantum chaotic many-body systems present intriguing challenges in quantum statistical mechanics \cite{kh,liusuh,kaufman2016,asplund2015,sekino2008,maldacena2016,aleiner2016,roberts2016,casini,hosur2016,ha,nahum,ms,mezeiholography,gu2016,luitz,nahum3,keyserlingk,nahum2,amosandreajohn}.
We consider here the bipartite entanglement, which may be quantified by the von Neumann or Renyi entanglement entropies.  
This is a property of a given  state or a given operator,
together with a  chosen ``cut'' that divides the system in to two parts.  
In integrable models, spreading quasiparticles provide a heuristic picture for entanglement growth \cite{CCreview}, but in chaotic models 
it may be more useful to think in terms of  
local entanglement ``production'', rather than entanglement ``spreading''. 
If the system is quantum chaotic and the entanglement across a particular cut is less than the maximal value it approaches at equilibrium, then the system's dynamics will generically produce additional entanglement across that cut, at a rate that is constrained by the entanglement at nearby cuts \cite{nahum}.   We consider systems with only short-range interactions, so the dynamics is local in that sense.  

In this paper we explore the ``hydrodynamics" of entanglement production \cite{nahum}.  We discuss both  the entanglement of pure quantum states and the entanglement of quantum operators.
Operators can be viewed as pure states in a doubled Hilbert space (``bra'' and ``ket''), 
so the definitions of von Neumann and Renyi entropies carry over directly from states to operators \cite{zanardi,prosenpizorn,prosenpizorn2,luitz,dubail}. The doubling of the Hilbert space means that under Heisenberg evolution operators generate entanglement at up to twice the rate for  the corresponding states.

The general picture is simplest for one-dimensional systems with one cut at position $x$.  The entanglement entropy for a given 
state as a function of the time and the position of the cut is $S(x,t)$.  The rate of entanglement entropy production at $x$ is, to leading order in a coarse-grained limit, set by a system-specific function, $\Gamma(\partial S/\partial x)$, of the spatial derivative of $S$.  As we will show below, this function $\Gamma$ encodes various aspects of the entanglement and operator dynamics, including both the ``entanglement speed'' $v_E$ and the ``butterfly speed'' $v_B$.

This picture has a dual ``spacetime'' interpretation 
in which the entanglement is mapped to the ``energy'' of a coarse-grained  curve, or in higher dimensions a membrane, which transects the spacetime patch \cite{nahum}.  In the scaling limit, this curve has a well-defined geometry that is determined by a ``line tension'' function $\lt(v)$ which depends on the local velocity of the curve. This function is related to the  entanglement production rate $\Gamma$ by a  Legendre transformation.
The entanglement line tension $\lt(v)$ is in general model dependent,  but we argue it satisfies various constraints.  We show how it may be obtained numerically.
Heuristically, $\lt(v)$ can be thought of as the appropriate coarse-grained ``cost'' function for a ``minimal cut'' through a unitary circuit generating the dynamics. In any tensor network, the length of the ``minimal cut'' separating two regions gives an upper bound on the entanglement \cite{swingleentanglementrenormalization, casini,pastawski,hayden}.
While this heuristic becomes precise in certain limits \cite{nahum}, in general  the coarse-grained minimal curve cannot be simply identified with a cut through a microscopic circuit.

The operator entanglement is a tool for quantifying the structure of  an operator in a basis-independent manner.
In 1D, the entanglement across spatial cuts is related to the cost of storing the operator in a  matrix-product-operator representation, just as the state entanglement is related to the cost of a matrix-product-state representation.
We discuss in detail the case of an initially local operator spreading out under Heisenberg evolution.
Contrary to the naive guess, we find that a spreading operator is far from being fully entangled within the region to which it has spread. 
Therefore a spreading operator is structurally very different to a generic random operator of the same spatial footprint. (This suppression of the operator entanglement is not however as strong as in certain integrable systems \cite{prosenpizorn,prosenpizorn2,dubail}, where a spreading operator can entangle sublinearly with time or not at all.) We give a scaling picture for the entanglement profile $S(x,t)$ of a spreading operator, which in 1D resembles an expanding pyramid.

The specific Hamiltonian model that we have used for exploring and testing these scaling pictures is the
quantum chaotic Ising spin chain with longitudinal and transverse fields:
\begin{equation}
H=\sum_{i=1}^{L-1} Z_i Z_{i+1} + h\sum_{i=1}^L Z_i + g\sum_{i=1}^L X_i ~,
    \label{eq:ham}
\end{equation}
where $X_i$ and $Z_i$ are the Pauli operators for the spin-1/2 at site $i$, $h=0.5$ and $g=-1.05$ (these choices follow Refs.~\cite{ms,mcb}).
We also use some results from random unitary circuits \cite{nahum,nahum3,keyserlingk,nahum2,amosandreajohn, zhounahum}.
{\blue Entanglement is measured in bits in all plots.}

\tableofcontents

\section{Scaling picture}
\label{scaling_picture_section}
\subsection{General features}

Let us begin with 1D, where we can motivate the minimal surface picture by considering a dynamical equation for the entanglement which may be more intuitive.  Consider a ``generic'' nonequilibrium pure state $|\psi\rangle$ of a finite isolated system; this state has entanglement well below thermal equilibrium and has been becoming more entangled under the system's quantum-chaotic local unitary dynamics.  Let $S(x,t)$ be the bipartite von Neumann entanglement entropy of that state across a cut at position $x$ at time $t$.  Then we will assume that the leading coarse-grained behavior of the local rate of increase of this entanglement entropy 
is determined by an entropy production rate $\Gamma(s)$, which is a function of the local gradient $s$ of the entanglement:
\begin{equation}
    \frac{\partial S}{\partial t} = \seq \, \Gamma\lf \frac{\partial S}{\partial x}\ri.
    \label{eq:state}
\end{equation}
We have  extracted a factor of $\seq$, the  entropy density of the state to which the system is equilibrating.  

The entropy production rate $\Gamma$ vanishes at equilibrium. If the equilibrium state has entropy density $\seq$, then the ``profile'' of its entanglement for a system of length $L$ is the pyramid 
\be
S(x,t)=\seq \, {\rm min}\{x, L-x\},
\ee
with slope $|\partial S/\partial x|= \seq$.  Thus $\Gamma(s)$ is positive when ${s=\partial S/\partial x}$ is between $-\seq$ and $+\seq$, since in this interval the state is not maximally entangled and the chaotic dynamics will generate additional entanglement, while $\Gamma(-\seq)=\Gamma(+\seq)=0$.  This function $\Gamma(s)$ is model-dependent and encodes not only the rate of entanglement growth but also some information about ``light-cone'' effects in correlation functions: the derivative $-\seq \Gamma'(\seq)$ is equal to $v_B$, the ``butterfly'' speed, which is the effective Lieb-Robinson \cite{lr} speed
 governing the spreading of operators, as we will discuss below.  The ``entanglement speed'' $v_E$, which sets the rate of entanglement growth for an initially unentangled state, is given by $v_E = \Gamma(0)$.

If the model has conserved densities, the entanglement dynamics will also be coupled to the transport of these densities, with $\seq$ and $\Gamma(s)$ depending on the local densities; for simplicity, here we assume that the coarse-grained spatial distribution of any such conserved densities is uniform, so the above equation (\ref{eq:state}) suffices.  In the presence of static spatial inhomogeneities \cite{nahum2},  $\Gamma$ also depends directly on the position $x$, but here we assume that the Hamiltonian  producing the dynamics is statistically spatially uniform away from the ends of the spin chain.  Finally, we  assume the dynamics is quantum chaotic, so we are not considering  integrable systems.  But apart from this we are discussing unitary time evolution rather generally, so the Hamiltonian $H$ may be time-independent, or it may be periodic in time and thus realize a Floquet time-evolution, or it may be a random function of time drawn from some ensemble, as in Ref. \onlinecite{nahum}.  In the two latter cases we must coarse-grain some in time to not ``see'' the short-time variations due to the time-dependence of $H(t)$. 

As applied to general nonintegrable systems, this picture is a conjecture which is supported by exact results for random circuits as well as numerical finite-size scaling analyses of entanglement saturation \cite{nahum,nahum2} including those presented here. In the spirit of thermodynamics, we assume that this coarse-grained description holds  on long length and time scales for ``generic'' initial states.  As with the thermodynamics of closed systems, it is possible to find atypical initial states for which it does not apply (Sec.~\ref{tracesection}).

\begin{figure}[t]
\centering{
\includegraphics[width=0.75\linewidth]{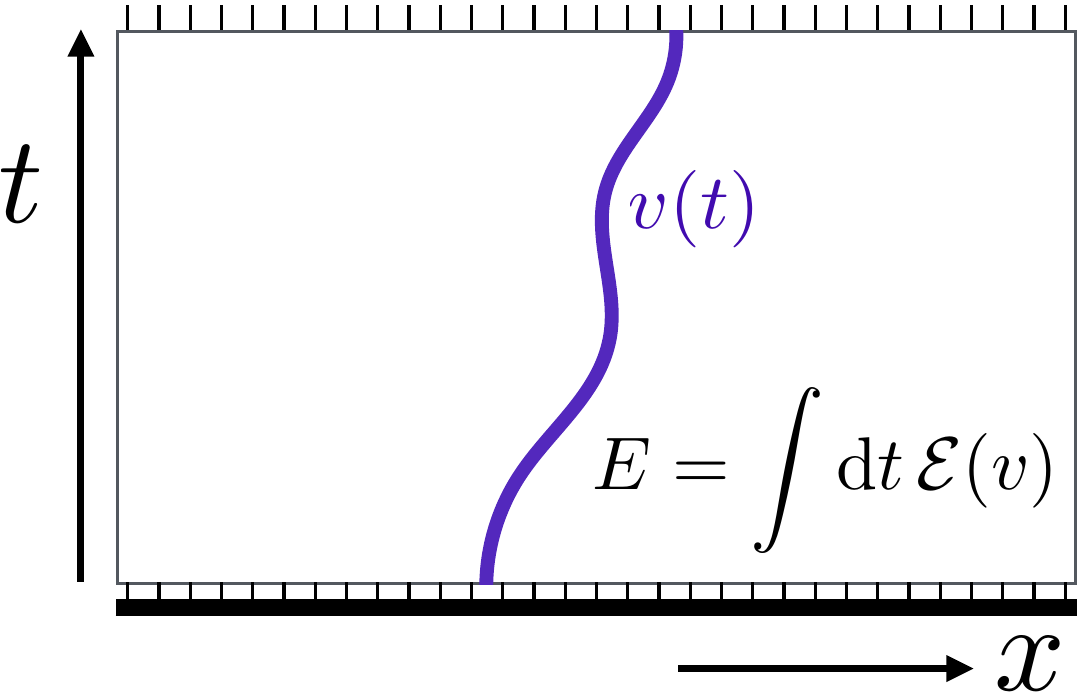}
}
\caption{
Minimal curve picture in 1+1D.  At each point in time the directed curve can be assigned a velocity $v(t)$.  Its entanglement ``energy'' is the integral of a velocity-dependent ``line tension'', plus a possible contribution from the initial state; see Eq.~\ref{spacetimeformula}.
}  \label{fig:minimalcurve}
\end{figure}

This picture in terms of a growth rate $\Gamma(s)$ is mathematically equivalent to a spacetime picture in terms of a coarse-grained minimal curve, where the crucial data is a function  $\lt(v)$ encoding the ``line tension'' of this curve \cite{nahum}.  This line tension is a coarse-grained measure of the entanglement across a spacetime cut through the unitary operator that generates the dynamics, as we discuss below.
In some limits \cite{nahum}, the minimal curve can be related to the idea of a minimal cut through a tensor network \cite{swingleentanglementrenormalization, casini,pastawski,hayden}, which gives a microscopic upper bound on the entanglement in any tensor network or unitary circuit.

The unitary evolution takes place in a spacetime patch of spatial extent $L$ and temporal extent $t$. We will consider directed curves which pass from the spacetime point $(x,t)$ at the final time either to $(y,0)$ at the initial time or to a point 
on the spatial boundary for systems with open ends.  Consider first the limit of an infinite chain, so that only the first option is allowed. 

A curve in spacetime has a velocity $v=dx/dt$ (Fig.~\ref{fig:minimalcurve}). We define a velocity-dependent line tension, $\lt(v)$, for such curves, again with dimensions of velocity.  The ``energy''  of a section of the curve of duration $\delta t$ and velocity $v$ is defined as $\seq\, \lt(v) \delta t$, which is dimensionless, and is a measure of the coarse-grained entanglement across this section of the curve.  In order to compute $S(x,t)$, we consider all directed curves which travel from spacetime point $(x,t)$ 
to an arbitrary position $(y,0)$ at the initial time.
Such a curve is assigned an energy which is the integral of the line tension energy along the curve, together with a piece $S(y,0)$ from the entanglement of the initial state: see Fig.~\ref{fig:minimalcurve}.  The entanglement is given by the energy of the minimal-energy such curve.  In the present setting this curve is a straight line with some constant velocity $v=(x-y)/t$, so that
\be\label{spacetimeformula}
S\lf x,t\ri = \min_y \bigg(
 \, t ~\seq \, \lt \lf \f{x-y}{t}  \ri + S\lf y, 0 \ri 
\bigg).
\ee
In cases where the dynamics is inhomogeneous in space \cite{nahum2} or in time,
$\lt(v)$ will acquire an additional explicit dependence on $x$ or $t$. The minimal curve will then no longer be a straight line.

\begin{figure}[t]
\centering{
\includegraphics[width=\linewidth]{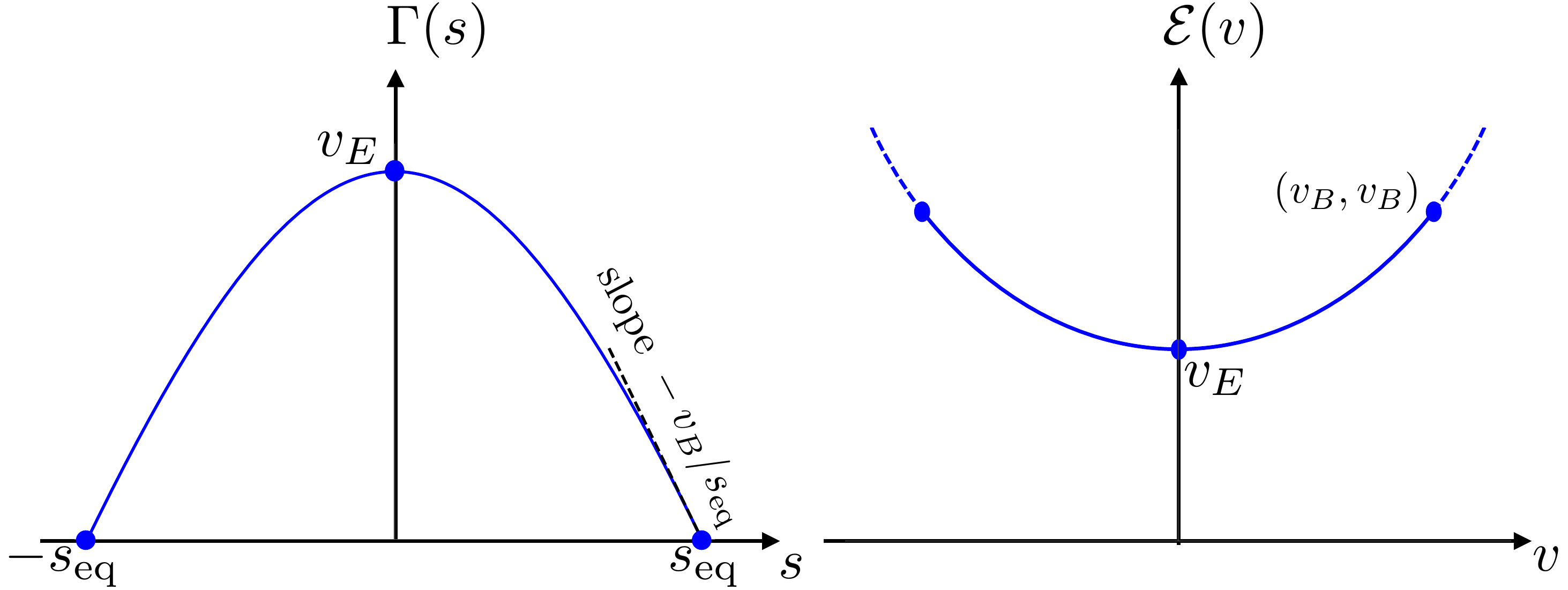}
}
\caption{
Schematic: entanglement growth functions $\Gamma$ and $\lt$ (from the finite $q$ random circuit example in Eqs.~\ref{regularexample},~\ref{regularexamplegamma}).
}  \label{egammafig}
\end{figure}

The equivalence between (\ref{eq:state}) and (\ref{spacetimeformula}) is seen by differentiating (\ref{spacetimeformula}) with respect to $t$. We obtain Eq.~\ref{eq:state} with 
\be\label{legendre_transform}
\Gamma(s) =  \min_v \bigg( 
  \lt (v)  -   \f{vs}{\seq}    \,
\bigg).
\ee
$\lt(v)$ is given in terms of $\Gamma(s)$ by the inverse Legendre transformation,
\be
\label{eq:inverselegendre}
\lt(v) = \max_s \left(\Gamma(s) +  \frac{vs}{\seq}  \right).
\ee
See Fig.~\ref{egammafig}.

In the above discussion, the minimal curve arose as a mathematical construct to solve Eq.~\ref{eq:state}. However the picture of a minimal curve or (in higher dimensions) surface with a coarse-grained surface tension $\lt(v)$ can be motivated independently, and it will often be useful to think of the curve or surface as the primary object.

The crucial feature of the minimal surface, which we will rely on in the following,  is an effective \textit{locality} of its coarse-grained description. The ``energy'' of the curve is simply an integral of the velocity-dependent line tension along the length of the curve.  This is a nontrivial assumption which certainly cannot hold for integrable systems, where a very different quasiparticle picture provides a useful description of entanglement growth \cite{CCreview}.  But  we conjecture  that this locality property is an emergent property of the dynamics of generic \textit{nonintegrable} systems on large scales \cite{nahum}.

For now we  assume spatial reflection symmetry ${\Gamma(s)= \Gamma(-s)}$ (this may be relaxed, see  Sec.~\ref{general_features}).
Eq.~\ref{eq:inverselegendre} defines $\lt(v)$  in the range ${|v|\leq v_\text{max}}$, where ${v_\text{max}=-\seq \Gamma'(\seq)}$. By basic properties of the Legendre transformation, and assuming $\Gamma$ to be analytic in the range $(-\seq, \seq)$, we have $\lt(v_\text{max}) = v_\text{max}$ and $\lt'(v_\text{max}) = 1$.
We will argue in Sec.~\ref{general_features} that $v_\text{max}$ is finite in models with local interactions, and that in general  $v_\text{max}$ is equal to $v_B$, the speed at which operators spread. 
 This argument involves an assumption that different natural measures of the ``size'' of a spreading operator 
are governed by the same growth speed $v_B$. 
If this assumption failed, then the speed $v_\text{max}$ 
relevant to $\lt$ and $\Gamma$ could correspond to a different ``operator spreading speed'' to the speed $v_B$ defined by the out-of-time-order correlator.
For  higher Renyi entropies, and for the von Neumann entropy in a certain limit,  nontrivial analytical checks on the identity $v_\text{max}=v_B$ are possible (Sec.~\ref{randomcircuitexamples}). We will also give a numerical consistency check.

We therefore have the important basic constraints on the line tension
\ba
\label{eq:vbconstraint}
\lt(v_B) &= v_B, & \lt'(v_B) & =1, &
&\text{with $\lt(v) \leq |v|$},
\end{align}  
together with the convexity condition  (Sec.~\ref{general_features})
\be\label{eq:convexity}
\lt''(v)\geq 0.
\ee
As a result of (\ref{eq:vbconstraint}), only curves with velocity less than or equal to ${v_B}$ are ever required for the minimization in Eq.~\ref{spacetimeformula}. It will suffice to consider only such curves. Interestingly however, the explicit random circuit calculation reviewed in Sec.~\ref{randomcircuitexamples} shows that there is a sense in which $\lt(v)$  remains well-defined (at least for the higher Renyi entropies) for speeds greater than $v_B$ but less than the strict causal light cone speed, if one exists.

\begin{figure}[t]
\centering{
\includegraphics[width=0.94\linewidth]{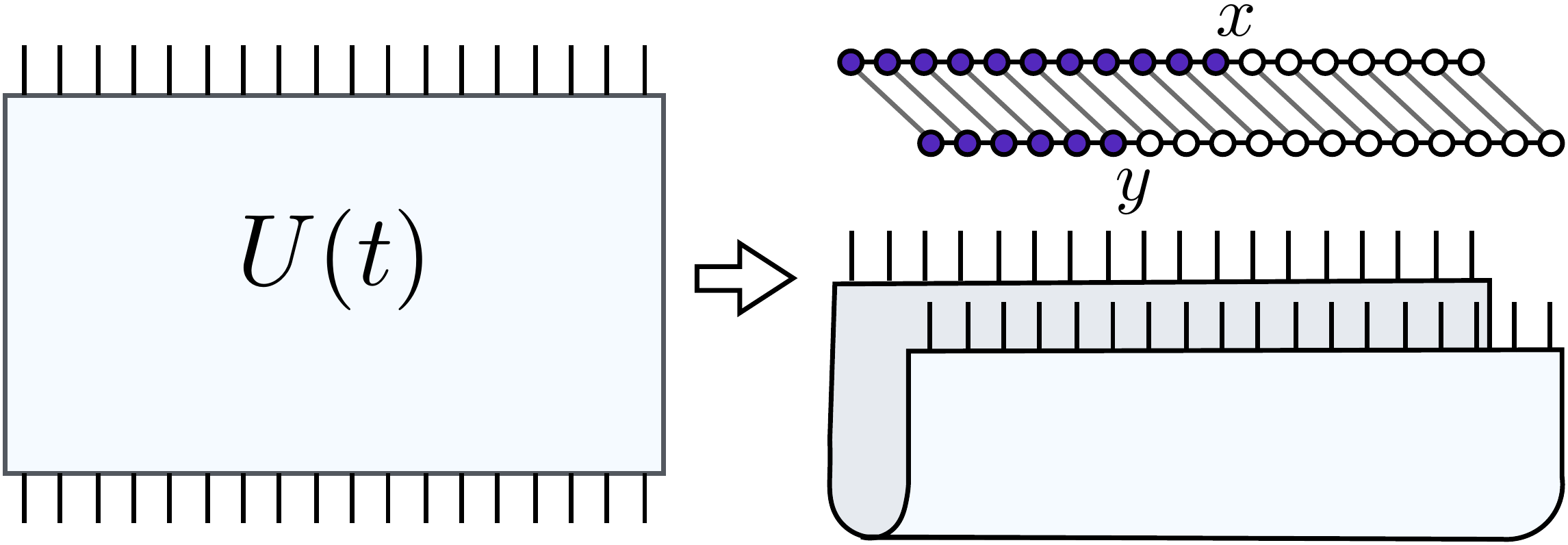}
}
\caption{
Entanglement of the time evolution operator.  The entanglement of an operator acting on $L$ spins, for example the time evolution operator (left), is calculated by treating it as a  \textit{state} on $2L$ spins  (right). $S_U(x,y,t)$ denotes the entanglement of the subsystem consisting of the shaded spins in the upper right figure. 
}  \label{fig:Uentanglement}
\end{figure}

The right hand side of (\ref{spacetimeformula}) has two parts, one coming from the initial state, and one which is independent of the initial state. At infinite temperature we identify this second part  with the \textit{operator entanglement} of the unitary time evolution operator, $U(t)$, which
advances time from zero to $t$ \cite{zanardi,luitz,dubail}. (We expect that a similar identification also holds at finite temperature, for an ``effective'' time evolution operator: this effective time evolution operator acts in the corresponding Hilbert space of lower-energy states, whose effective local dimension $q_\text{eff}$ is determined by the thermal entropy density, $\seq = \log q_\text{eff}$.)

Recall that the operator entanglement \cite{zanardi,prosenpizorn,prosenpizorn2,luitz,dubail} is defined by treating the operator as a state in a doubled Hilbert space, with the two sets of ``spins'' corresponding to the row and column indices of the operator respectively: see the cartoon in Fig.~\ref{fig:Uentanglement}.   Visually, if we think of $U(t)$ as a matrix product operator, with ``legs'' at the top representing the row indices and legs at the bottom representing the column indices, then the mapping to a state simply means treating this object as a matrix product state in which both the upper and lower legs are  physical spin indices. (We review the definition of operator entanglement in more detail in Sec.~\ref{operator_entanglement_definition}.)

Let  $S_{U}(x,y,t)$ denote the entanglement of $U(t)$, for a cut that makes a ``subsystem'' which includes all the row spins to the left of $x$ and all the column spins to the left of $y$: see Fig.~\ref{fig:Uentanglement}, upper right.  Note that at time zero, $U(0)$ is simply the identity: this corresponds to a  state in which each row spin is maximally entangled with the corresponding column spin, but in which spins at distinct spatial sites are not entangled. This means that ${S_U(x,y,0) = \seq |x-y|}$, where $\seq$ is the logarithm of the Hilbert space dimension (since we are temporarily restricting to infinite temperature).

\begin{figure}[t]
\centering{
\includegraphics[width=0.7\linewidth]{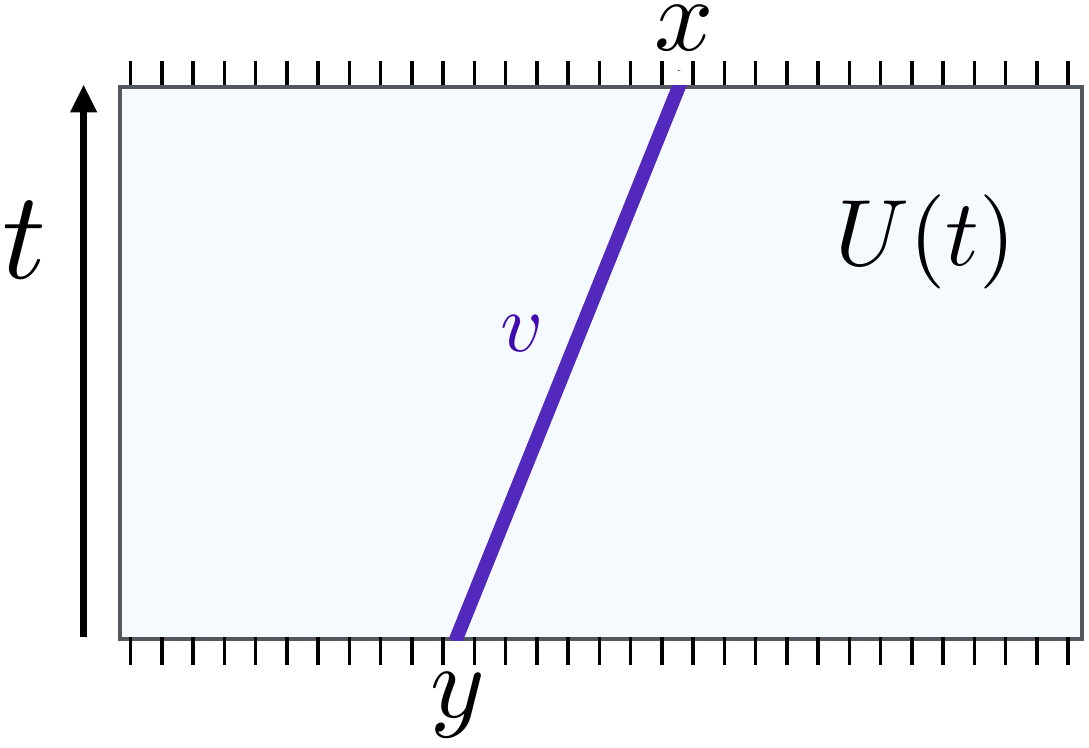}
}
\caption{The unitary entanglement $S_U(x,y,t)$ is proportional to the line tension $\lt(v)$ for a cut with $v=(x-y)/t$ (when $|x-y|\leq v_B t$, and neglecting boundary effects).}
 \label{fig:unitarycut}
\end{figure}

\begin{figure}[t]
\centering{
\includegraphics[width=0.95\linewidth]{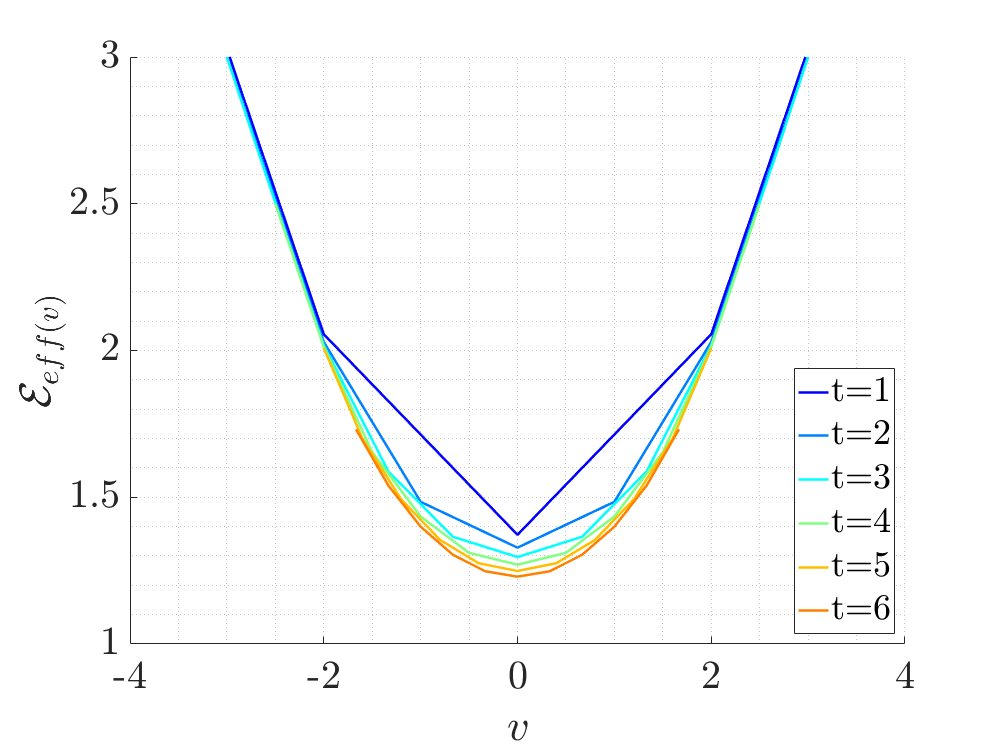}
}
\caption{
Numerical determination of the entanglement line tension of the nonintegrable Ising model. $\lt_\text{eff}$ is defined in Eq.~\ref{eeffdefn} and is expected to converge as in Eq.~\ref{eq:eveff} at late times. This data is for a system of size $L=12$. Here ${v\equiv |x-y|/t}$, and $(x+y)/2$ is at the centre of the chain to minimize boundary effects (hence $|x-y|$ is even). See also Fig.~\ref{fig:isingev2}.
}  \label{fig:isingev}
\end{figure}

\begin{figure}[b]
\centering{
\includegraphics[width=0.49\linewidth]{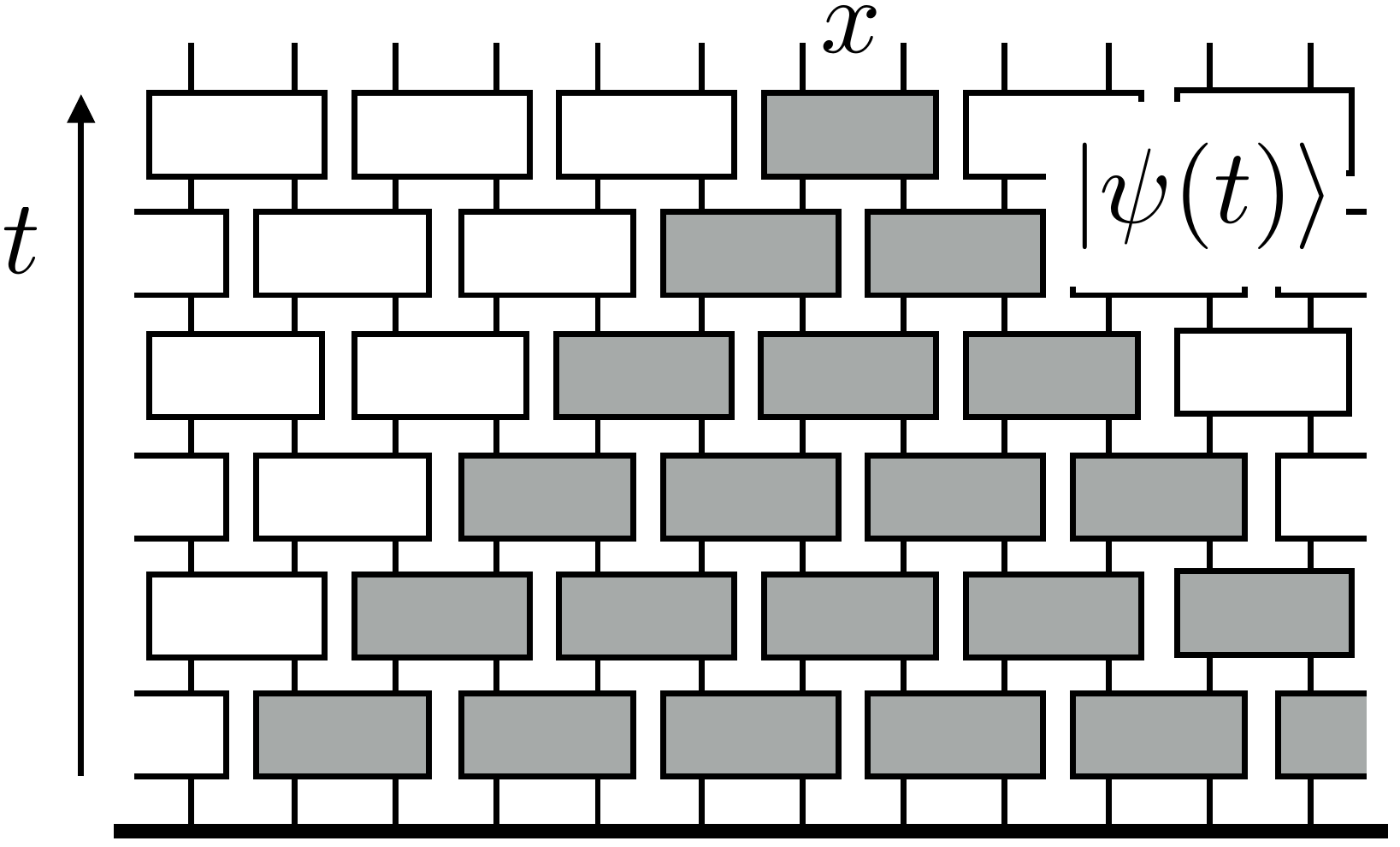}
\includegraphics[width=0.49\linewidth]{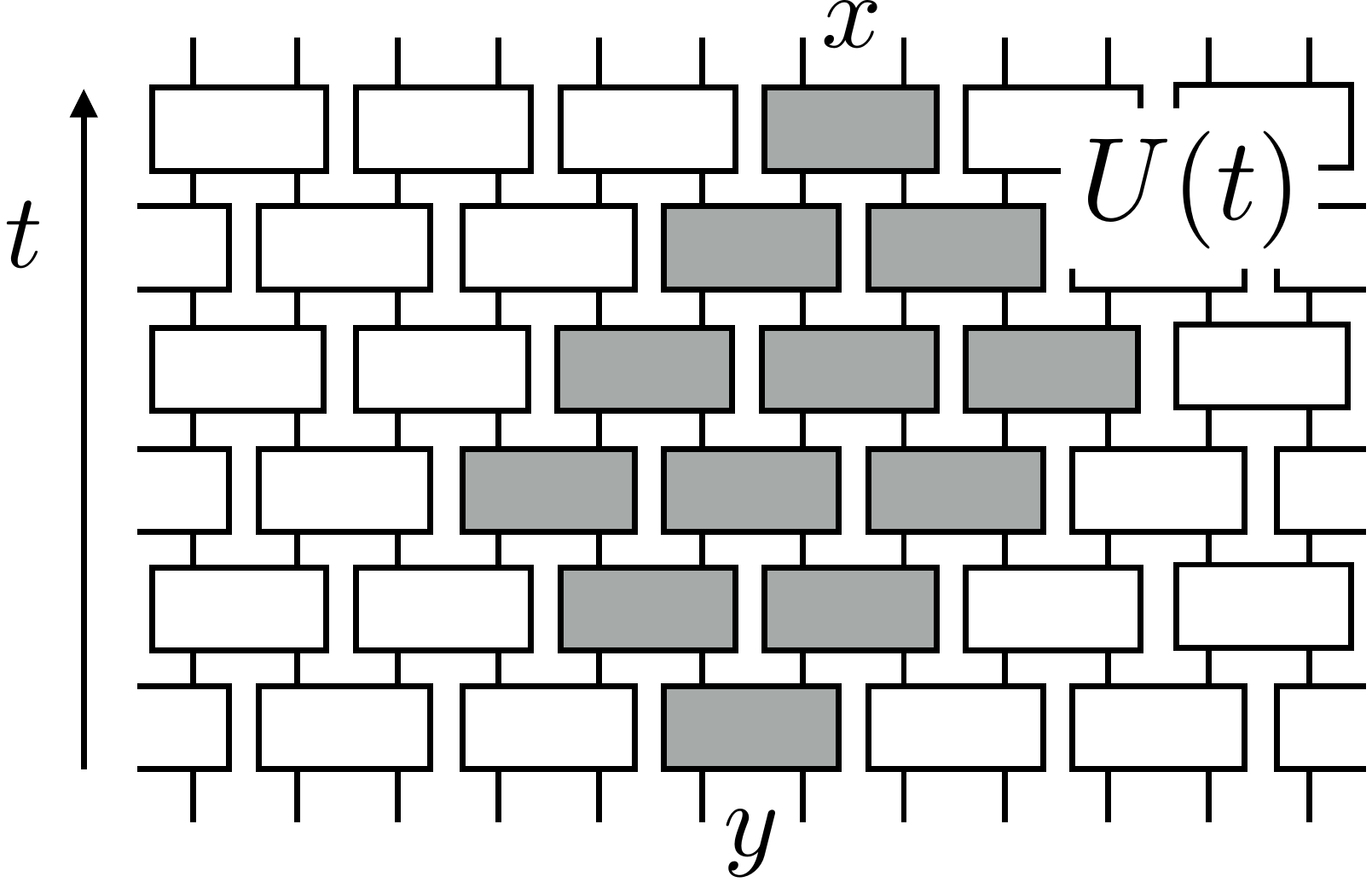}
}
\caption{A unitary circuit has a ``naive'' lightcone speed set by the geometry of the circuit. The state entanglement  $S(x,t)$ (left)  is unaffected if the unitaries outside the shaded lightcone are removed, i.e. replaced with the identity. Similarly, the operator entanglement $S_U(x,y,t)$ of $U(t)$ is unaffected if the  unitaries outside the shaded intersection of the past lightcone of $x$ and forward lightcone of $y$ are removed. The leading order entanglement dynamics postulated here is consistent with a lightcone effect in which the naive lightcone speed is replaced by $v_B$. See Eq.~\ref{S_U_for_large_x}: when $|x-y|\geq v_Bt$, the effective lightcone is empty and $S_U(x,y,t)$ is equal to its $t=0$ value.
}
 \label{fig:lightcones}
\end{figure}

 In the scaling limit, $S_{U}(x,y,t)$ is given by the energy of a cut that runs from $(x,t)$ at the final time of the spacetime patch to $(y,0)$ at the initial time:
\ba\label{eq:SU1}
S_{U}(x,y,t) & = t\times \seq\,\lt\lf \f{x-y}{t}\ri & 
& \text{for $|x-y|\leq v_B t$}.
\end{align}
When $|x-y|=v_B t$, the above formula matches the $t=0$ result (by Eq.~\ref{eq:vbconstraint}).
That is, the change of $S_U(x,y,t)$ 
from its $t=0$ value is exponentially small \cite{lr} in $t$ and thus can be ignored in the scaling limit as long as $|x-y|/t$ exceeds $v_B$:
\ba\label{S_U_for_large_x}
S_{U}(x,y,t) & = \seq |x-y| & 
& \text{for $|x-y| \geq  v_B t$}.
\end{align}
We may interpret Eq.~\ref{S_U_for_large_x} in the spacetime picture as  a minimal cut with a section at speed $v_B$, together with a horizontal section at the lower boundary which costs an energy equal to $\seq$ multiplied by its length.

Eq.~\ref{S_U_for_large_x} is also consistent with the expectation that the dynamics in some regions of the spacetime patch cannot affect the entanglement across a given cut, due to effective causality constraints \cite{ms}, so that the circuit can be ``truncated'' (i.e. the local unitaries in those regions, in  a quantum circuit picture, can be replaced with identities).
This is simplest to motivate using  a discrete time quantum circuit, which has a strict causal light cone speed: see Fig.~\ref{fig:lightcones}. This speed defines, for any entanglement cut, a causal cone outside which the unitaries cannot have any effect on the entanglement. These unitaries can be discarded. 
Eq.~\ref{S_U_for_large_x} is consistent with the naive picture in which the (in general) smaller speed $v_B$ defines an \textit{effective} causal cone for the leading order dynamics.
When  $|x-y|\geq v_B t$ this effective causality constraint allows all the unitaries in the circuit to be removed, so that in the scaling limit $S_U/|x-y|$ is unchanged from its $t=0$ value when $|x-y|/t=v \geq v_B$. 

\begin{figure}[t]
\centering{
\includegraphics[width=0.97\linewidth]{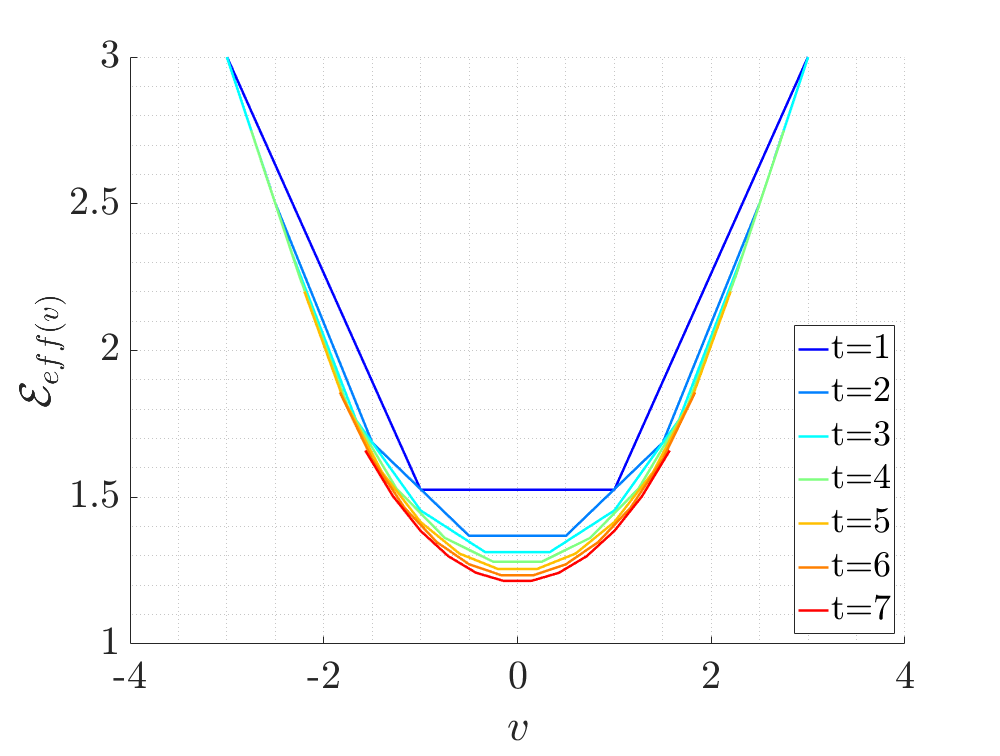}
}
\caption{
As in Fig.~\ref{fig:isingev} but for $L=13$ and odd values of $|x-y|$. The convergence of $\lt(0)$ is relatively slow --- we estimate the asymptotic value to be in the range $(0.95,1.1)$.
}  \label{fig:isingev2}
\end{figure}

The entanglement $S_U(x,y,t)$ may be viewed as the central object in the minimal curve picture (which generalizes to a minimal surface in higher dimensions as we discuss below).  This quantity also provides a useful means of determining $\lt(v)$ numerically.

We use the nonintegrable Ising model (\ref{eq:ham}) to illustrate a numerical procedure for obtaining $\lt(v)$.  For simplicity, imagine first a system that is spatially infinite, so that finite size effects can be neglected. Finite \textit{time} effects must still be taken into account. For a given $t$ we define an estimate of the line tension via
\be\label{eeffdefn}
\lt_\text{eff} (v)  \equiv \f{1}{\seq\, t} \times S_U \lf - \f{v t}{2} , \f{v t}{2} , t \ri.
\ee
By Eqs.~\ref{eq:SU1},~\ref{S_U_for_large_x}, this converges at large $t$ to 
\be\label{eq:eveff}
\lt_\text{eff}(v) \rightarrow
\left\{
\begin{array}{cc}
 \lt(v)& \text{for $|v|\leq v_B$\phantom{.}} \vspace{0.5mm}  \\ 
 |v| & \text{for $|v|\geq v_B$.}
\end{array}
\right.
\ee
In Figs.~\ref{fig:isingev},~\ref{fig:isingev2} we show numerical results for $\lt_\text{eff}$ using, respectively, even values of $|x-y|$ in a system of size $L=12$, and odd values in a system of size $L=13$. The maximum times ($t_\text{max} = 6$ and $7$ respectively) are limited by finite $L$ effects, which become strong at later times, as shown in  Appendix~\ref{app:SUdata}.  At late times, the preferred minimal cut configuration travels to a spatial boundary instead of resembling Fig.~\ref{fig:unitarycut}.

It is important to note that finite time effects are relatively strong in Figs.~\ref{fig:isingev} and \ref{fig:isingev2} ---  this can be understood in terms of subleading corrections to Eq.~\ref{eq:state} which we discuss in Sec.~\ref{higher_gradient}. As $t\rightarrow\infty$ the minimum of the curve should converge to $v_E$, which we estimate in Sec.~\ref{higher_gradient} to be in the range 
\be
v_E \in (0.95, 1.1)
\ee
(this estimate is consistent with \cite{ms}). The data for $t\sim 7$ are still above this range.

Nevertheless, Fig.~\ref{fig:isingev} nicely illustrates the key features of the line tension. The data is consistent with gradual convergence to a well defined $t\rightarrow \infty$ form which is analytic for $|v|< v_B$ and equal to $|v|$ for $|v|\geq v_B$, in accord with Eq.~\ref{eq:eveff}. Looking for the value of $v$ in Fig.~\ref{fig:isingev} where $\lt(v)=v$ indicates $v_B \sim 2$. This is roughly consistent with our independent numerical determination of $v_B$ from an analysis of a spreading operator in Appendix.~\ref{app:extrapolationsspr}, which gives
\be
v_B \simeq 1.8.
\ee
This is close to estimates in \cite{ms}.

Another basic consistency check on the scaling theory presented here is that measuring the entanglement $S_U(x,x,t)$ of the unitary, and measuring the entanglement growth rate for an initially unentangled state, should yield the same value for $v_E=\lt(0)$. Numerical results are consistent with this, as we discuss in Sec.~\ref{higher_gradient}. The nature of the finite $t$ corrections is different in the two cases, and larger for $S_U$. 

\begin{figure}[b]
\centering{
\includegraphics[width=\linewidth]{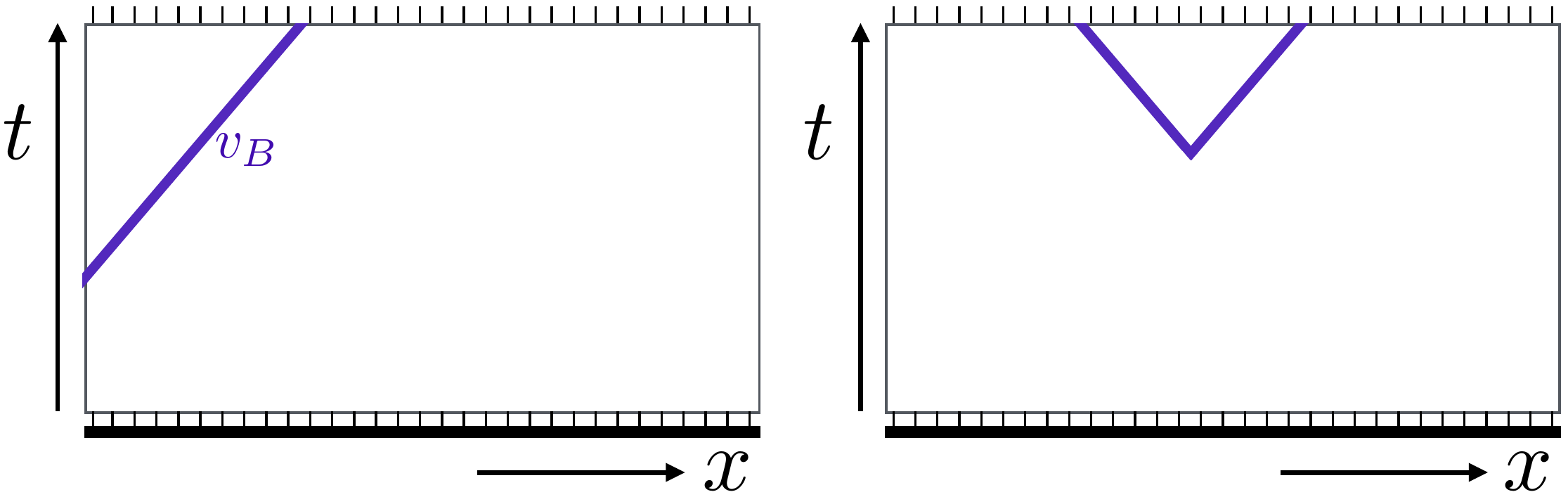}
}
\caption{
Minimal cut configurations at late time, after saturation of the entanglement, for a set of spins on the left of the chain (left) and a finite interval in a large system (right).
}  \label{fig:saturation}
\end{figure}

So far we have discussed infinite systems. In a finite system the entanglement $S(x,t)$ of a state evolving after a quench eventually saturates.  In this picture this occurs when a  minimal curve that exits via the boundary of the system (and travels at velocity $\pm v_B$) has lower energy than one which reaches the $t=0$ boundary. This crossover is discontinuous in the scaling limit which we are considering here, and for a quench from an unentangled initial state in 1D it gives a piecewise linear scaling form for $S(x,t)$ \cite{nahum}.  Similarly, for a finite interval saturation occurs when the minimal curve no longer reaches the $t=0$ boundary.  This is illustrated in Figs.~\ref{fig:saturation}.

\subsection{Examples of $\Gamma$ and $\lt$ from random circuits}
\label{randomcircuitexamples}

Random quantum circuits yield solvable models in which the entanglement growth functions $\lt$ and $\Gamma$ can be computed explicitly.\cite{nahum, nahum3, zhounahum} These models describe spin chains with $q$-state spins. The models have no conserved quantities, so the equilibrium entropy density $\seq = \log q$ is set simply by the local Hilbert space dimension.  In one simple nontrivial model\cite{nahum} the entanglement production rate and line tension in the large $q$ limit have simple quadratic forms\footnote{The special case $\Gamma(0) = v_E =1/2$ is explained in \cite{nahum}. General $s$ is an immediate extension.}
\ba\label{randomstructureexample}
\Gamma(s) & = \f{1}{2} \lf 1 - \lf \f{s}{\seq}\ri^2 \ri, &
\lt(v) & =  \f{1}{2} \lf 1 + v^2 \ri.
\end{align}
We  have set the microscopic timescale of the dynamics and the lattice spacing to unity. The  `butterfly speed' in this model, governing the speed at which an initially local operator spreads out, is 
\be
v_B=1.
\ee
Note that $\lt(v_B) = v_B$ and $\lt'(v_B) = 1$ in accord with Eq.~\ref{eq:vbconstraint}.

This model has $q=\infty$. The dynamics consists of the application of random unitaries to each bond in a Poissonian fashion, at rate 1.  The corresponding quantum circuit has a random structure in spacetime.  In this strict $q=\infty$ limit, the minimal curve has a simple interpretation: it may be thought of as the cut through this random circuit which cuts the minimal number of bonds, and $\lt$ gives the number of bonds that are cut per unit ``length'' in the time direction.

For quantum circuits with {finite} local Hilbert space dimension $q$, the minimal curve can no longer be identified with a simple ``minimal cut'' at the lattice level. For example, the entanglement growth rate in general depends nontrivially on which Renyi entropy we consider once $q$ is finite\cite{zhounahum,amosandreajohn}.  

However, for random circuits with regular spatial structure\cite{zhounahum, nahum3, keyserlingk} the calculation of $\overline{e^{-S_2}}$, where the average is over the unitaries in the circuit, may be mapped to an effective statistical mechanics problem involving a directed ``polymer'' \cite{nahum3}. 
This picture may be extended  to the calculation of $\overline{S_2}$ using the replica trick, taking $q$ to be large but finite \cite{zhounahum}. After coarse-graining, the resulting polymer acquires a definite coarse-grained geometry, and becomes precisely the minimal curve discussed above. 

We point out here that the free energy of a polymer with a constrained slope determines the  line tension $\lt(v)$ for the coarse-grained minimal curve.
For $S_2$ (in general $\lt$ depends on the Renyi index; elsewhere in the paper we focus on the von Neumann entropy),\footnote{The corrections to this formula are at order $1/(q^8 \ln q)$ \cite{zhounahum}.}
\ba\label{regularexample}
\lt_2 (v) \simeq & \log_q  \f{q^2+1}{q} 
 +  \f{1+v}{2}  \log_q    \f{1+v}{2}   +    \f{1- v}{2}  \log_q    \f{1-v}{2}.
\end{align}
By Eq.~\ref{legendre_transform}, the corresponding growth rate is
\ba \label{regularexamplegamma}
\Gamma_2(s) & = \log_q \lf \f{q+q^{-1}}{q^{s/\seq}+ q^{-s/\seq}} \ri,
\end{align}
see Fig.~\ref{egammafig}. Finally, for this model\cite{nahum3, keyserlingk}
\be
v_B = \f{q^2-1}{q^2+1}.
\ee
Again note that $\lt_2(v_B) = v_B$ and $\lt_2'(v_B)=1$. When $q$ is finite the butterfly speed $v_B$ is smaller than the ``naive'' lightcone speed, which is unity. The latter is the bound on signal propagation which follows trivially from the geometry of the circuit in spacetime.  

Remarkably, recent work \cite{amosandreajohn} has shown that a Floquet model built from random unitaries (a circuit that is random in space but periodic in time) is solvable in the limit of large $q$. This leads to a domain wall picture for $\overline{e^{-S_2}}$ which at $q=\infty$ coincides with that discussed above. Therefore the expansion of Eq~\ref{regularexample} up to order $1/\ln q$ should also apply to the model of Ref.~\cite{amosandreajohn} (more precisely, this will be true up to an extremely long timescale when rare region effects come into play\footnote{We also assume here that averaging before/after the exponential will give the same result at leading order in $q$, as is the case in the fully random circuit.} \cite{nahum2}). The analytical results of Ref.~\cite{amosandreajohn}, in a model with time translation symmetry, are further support for the general validity of the minimal surface picture.

\subsection{General features of the line tension}
\label{general_features}

The above examples show that the functional form of the line tension depends on the model considered. (Universality may nevertheless arise in the low temperature limit, see Sec.~\ref{conclusions}.) 
In order for $\lt(v)$ to be a valid coarse-grained line tension it must satisfy  $\lt''(v)\geq 0$.  (If $\lt''(v_0)< 0$ for some $\lt(v_0)$, we can construct a path with coarse-grained slope $v_0$ whose coarse-grained energy density is smaller than $\lt(v_0)$, showing that $\lt(v)$ is not the correct coarse-grained line tension.) $\lt(v)$ must also satisfy the constraints in Eq.~\ref{eq:vbconstraint}, which we now discuss.

As noted in Sec.~\ref{scaling_picture_section}, the growth rate $\Gamma$ defines a maximal velocity $v_\text{max}$ via ${v_\text{max} = - \seq \Gamma'(\seq)}$. This is the velocity of the minimal curve when the initial entanglement gradient is maximal, $\partial S/\partial x = \seq$, and it is the maximal velocity of any minimal curve. Assuming $\Gamma$ is analytic, the fact that $\Gamma(\seq)=0$ implies Eq.~\ref{eq:vbconstraint} with $v_\text{max}$ in place of $v_B$.  We may argue that $v_\text{max} = v_B$ by considering the time-dependence of the entanglement in situations where the entanglement gradient is close to $\seq$, see below. The equality $v_\text{max} = v_B$ is also consistent with the heuristic picture, discussed above, of truncating the unitary circuit generating the dynamics. For example, it implies that in calculating the state entanglement $S(0, t_f)$ to leading order, we can truncate the spacetime patch to the region ${|x|<v_B(t_f-t)}$. For the regular quantum circuit in the limit $q\rightarrow \infty$, the butterfly speed $v_B$ becomes unity. In this limit the fact that we can delete the unitaries outside the lightcone without affecting the entanglement follows trivially from the circuit geometry, as shown in Fig.~\ref{fig:lightcones}. Above we have tested the relation $\lt(v_B)=v_B$ numerically in a realistic model.

\begin{figure}[t]
\centering{
\includegraphics[width=0.9\linewidth]{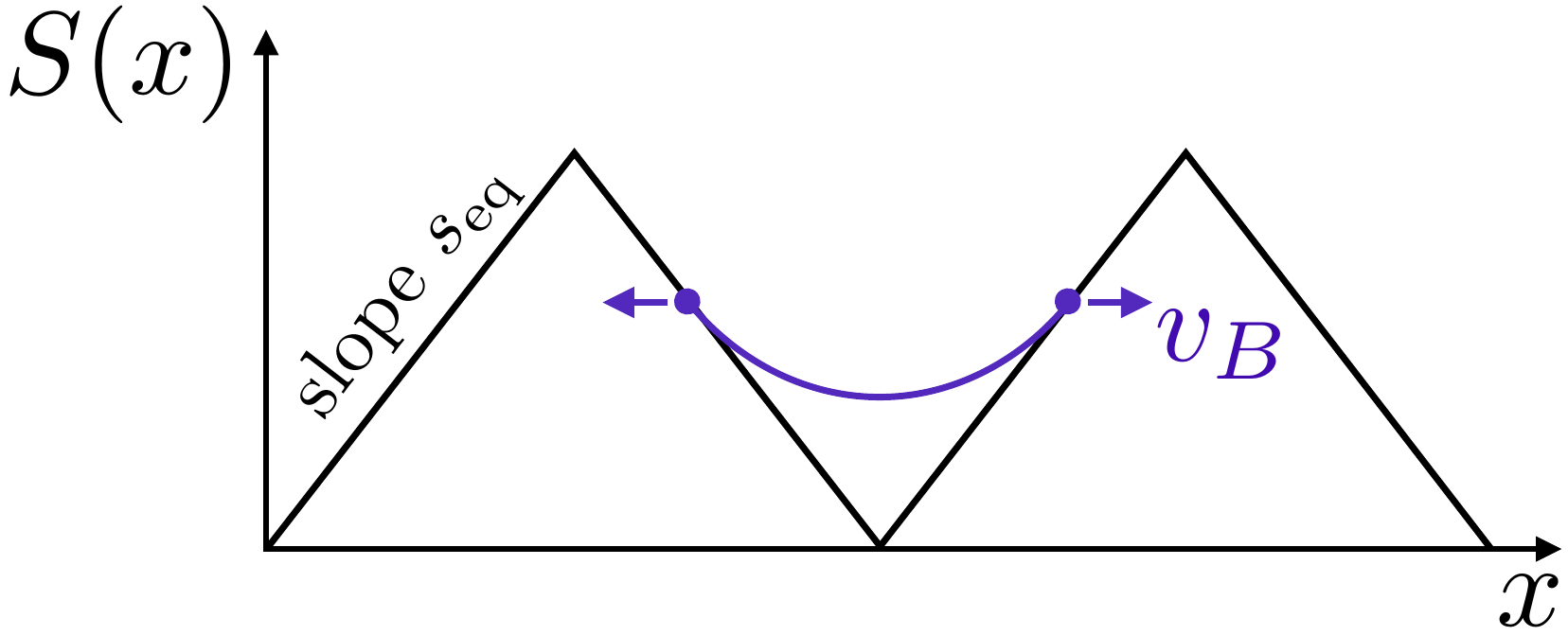}
}
\caption{
Separately thermalized half-chains yield the two-pyramid entanglement profile shown in black. At $t=0$ the two half-chains are connected.  As the subsystems entangle the region between the pyramids fills in (purple line).  The two points at which $S(x,t)$ departs from the initial profile move at speed $v_\text{max}$, which we argue is equal to $v_B$.
}  \label{fig:joiningprocess}
\end{figure}

The speed $v_\text{max}$ determines how fast ``features'' in the entanglement profile travel when $\partial S/\partial x$ is close to $\seq$.\footnote{$v(s) = - \seq \Gamma'(s)$ plays a similar role when the slope is  $\simeq s$.
For example, consider an initial linear entanglement profile ${S(x,0)= s x}$, and another initial profile ${\widetilde S(x,0) = s x+ \Delta(x,0)}$. To linear order in $\Delta=\tilde S-S$ we have $(\partial_t + v(s) \partial_x) \Delta (x,t)=0$. By the concavity of $\Gamma$, $v(s)$ is maximal for $s=\seq$. For the two random circuit examples discussed above we have respectively
$v(s)= \f{s}{\seq}$ and $v(s)= \f{q^{2 s/\seq}-1}{q^{2 s/\seq}+1}$.}  
For example,  consider an initial state of a chain in which the left and right halves are separately equilibrated, but the two halves are not entangled with each other. The initial entanglement profile is then a pair of adjacent pyramids with slope $\seq$, as shown in Fig.~\ref{fig:joiningprocess}. At $t=0$ the two subsystems are joined and begin to entangle, so that the region in between the pyramids fills in. It is straightforward to check using (\ref{eq:state}) or (\ref{spacetimeformula}) that the entanglement at a distance $x$ from the join first begins to grow at time $t = x/v_\text{max}$.
This is shown in Fig.~\ref{fig:joiningprocess} (where we have used $v_\text{max}=v_B$). This is the time required before operations in the vicinity of the origin can affect the entanglement (mutual information) between the regions to the left and right of $x$. Therefore $v_\text{max}$ encodes constraints due to causality on the time-dependence of the entanglement. This suggests that $v_\text{max}$, and therefore $\lt(v_B)$, should be identified with $v_B$. 

For a  more detailed heuristic argument we consider a modified version of the above protocol, where the two subsystems only interact for a finite time, with an evolution operator $U_\text{loc}$ acting on a finite region around the join, and then again evolve as separate systems. We may then write the final state in terms of the action of a ``time-evolved'' $U_\text{loc}$. This time-evolved operator grows at the operator spreading speed $v_B$. Making the assumption that this operator generates entanglement everywhere within its footprint when it is applied to separately thermalized subsystems yields $v_\text{max} = v_B$.\footnote{Take the initial state $\ket{\psi}=\ket{\psi_\text{left}}\otimes \ket{\psi_\text{right}}$ to have no entanglement between the two halves, but to be a ``generic'' thermalized state of the unjoined system, obtained for example by acting for a long time with the evolution operator $U_\text{sep}$ of the unjoined system on an appropriate initial state. $\ket{\psi}$ has a 2-pyramid entanglement profile with slopes $\seq$ for the pyramids. Next allow a period of evolution only in the vicinity of the cut with a time evolution operator $U_\text{loc}$ that acts on a patch of size $2R$, and finally evolve the two halves as separate systems with $U_\text{sep}(t)$. The final state $U_\text{sep}(t) U_\text{loc} \ket{\psi}$ may also be written $\widetilde U_\text{loc} \ket{\psi'}$, where $\widetilde U_\text{loc} \equiv U_\text{sep}(t) U_\text{loc} U_\text{sep}(t)^\dag$ and ${\ket{\psi'} = U_\text{sep} \ket{\psi}}$. We assume that $\widetilde U_\text{loc} \equiv U_\text{sep}(t) U_\text{loc} U_\text{sep}(t)^\dag$ is now effectively of extent $2 (R+ v_B t)$, i.e. that $U_\text{loc}$ grows at the butterfly speed when evolved with $U_\text{sep}$. Since $\ket{\psi}$ is a generic thermalized state with respect to $U_\text{sep}$, so is $\ket{\psi'}$, which also has the 2-pyramid entanglement profile. Since $\widetilde U_\text{loc}$ acts on all $x$ with $|x|\leq R+v_Bt$, we expect $S(x)$ to grow everywhere in that region when $\ket{\psi'}\rightarrow \widetilde U_\text{loc} \ket{\psi'}$. For consistency with the dynamics generated by $\Gamma$ this requires $v_\text{max}=v_B$, as stated above.} 

In this  argument we have assumed that the relevant ``size'' of the growing operator is the one dictated by the butterfly speed $v_B$ defined via the out-of-time-order correlator. 
In the future the relationship between different measures of the ``size'' of a spreading operator should be examined more carefully.

Note that for the second (``regular'') random circuit model described above, Eqs.~\ref{regularexample}--\ref{regularexamplegamma}, the limit $q\rightarrow \infty$ is pathological: $\Gamma(s)$ becomes a nonanalytic piecewise linear function and $\lt$ becomes flat. 
 These features occur whenever $v_E=v_B$, as a result of the constraint $\lt(v_B) = v_B$. (1D conformal field theories also have $v_E=v_B$, suggesting that entanglement generation in such CFTs is non-generic, even in irrational CFTs for which the quasiparticle picture does not apply.)  This nonanalyticity arises because in the strict $q=\infty$ limit the geometry of the minimal curve becomes ambiguous.\cite{nahum} This pathology is cured either by randomizing the structure of the circuit, as in the first example in Sec.~\ref{randomcircuitexamples}, or by making $q$ finite. 
 
The $q\rightarrow \infty$ limit also has the feature that $v_B$ coincides with the strict lightcone speed of unity set by the circuit geometry. 
At finite $q$ the explicit construction of the minimal curve in the random circuit, for the higher Renyi entropies, yields a finite $\lt_n(v)$  for all $|v|\leq 1$ , where $1$ is the strict light cone speed, which exceeds $v_B$. The part of the function with $v>v_B$  is irrelevant to the minimization in Eq.~\ref{spacetimeformula}, but it it is still physically meaningful (related for example to exponentially small outside-of-the-lightcone effects \cite{zhounahum,lv}).

In Ref.~\cite{ms}, two bounds on the growth rate of the entanglement entropy were proposed: one in terms of $v_E$,  and another based on causality with a light cone velocity $v_\text{LC}$. 
Assuming ${v_\text{LC}=v_B}$,  the minimal curve picture satisfies these bounds so long as ${\lt(v)\leq v_E + |v| (1-v_E/v_B)}$. This is  guaranteed by 
Eqs.~\ref{eq:vbconstraint},~\ref{eq:convexity}.
A theory whose minimal membrane has this piecewise linear $\lt(v)$ is equivalent  to a dynamics which 
saturates the bounds from Ref.~\cite{ms}.
 
So far we have assumed that the entanglement growth functions are inversion symmetric:  $\Gamma(s)=\Gamma(-s)$. This will be the case if the Hamiltonian has a spatial inversion symmetry (as for Eq.~\ref{eq:ham}) or a statistical inversion symmetry (as for the random circuits). But we may also consider 1D chaotic systems that are ``chiral'', in the sense that they have entanglement production dynamics that are not symmetric under spatial inversion, with differing butterfly speeds $v_B^+$ and $v_B^-$ for the right- and left-moving edges of a spreading operator and $\Gamma(s)\neq \Gamma(-s)$.\footnote{Random circuits built from ``staircase'' unitaries  \cite{nahum2} are chiral in this sense if  the densities of right and left staircases are not equal.} The above constraints are straightforwardly modified for this case.

\subsection{Higher dimensions}

The scaling picture above generalizes directly to higher dimensions.\cite{nahum}  In $d+1$ dimensions the curve generalizes to a $d$-dimensional surface.  Its local velocity $v$ is defined using the local tangent vector at the minimal angle to the time axis.  The ``energy'' is obtained by integrating the $v$-dependent local surface tension over the surface.  For a region bounded by a simple $(d-1)$-dimensional ``curve'' $C$, the state entanglement is:
\be\label{higher_d_S}
S(C,t)   =  \min \lf  S(C',0) + \, \seq \int \dd t' \dd^{d-1} x \,  \lt(v)
 \ri.
\ee
The minimization is over surface (membrane) shapes, and the integration is over the membrane.  The membrane terminates at $C$ on the $t'=t$ boundary and at the ``curve'' $C'$ on the $t'=0$ boundary. The discussion in the previous section carries over directly (so that again $\lt(v_B) = v_B$) as can be seen by considering a setup that is translationally invariant in all but one direction, so effectively 1D.  As in 1D, the second term on the right hand side of (\ref{higher_d_S}) can be interpreted as an entanglement of the unitary, which now  depends on the boundary curves $C$ and $C'$. 

The above equation may also be written in differential form, generalizing Eq.~\ref{eq:state}. For notational simplicity, consider the 2+1D case. We then have ${\partial S(C)/\partial t =\seq \int \dd l \, \Gamma (D(l))}$, where the integral is over the curve $C$, and $D(l)$ denotes the functional derivative of $S(C)$ with respect to a normal displacement $n(l)$  of the curve $C$ at position $l$: $D(l) = \delta S(C)/\delta n(l)$.

In writing Eq.~\ref{higher_d_S} we implied symmetry under spatial rotations and reflections.  But in a lattice model 
there is in general no reason to assume more symmetry for $\lt(\vec v)$ than that of the lattice point group, so $\lt$ will  depend on the orientation of $\vec v$ with respect to the lattice, and not only on $|\vec v|$.  This is consistent with the observation that in general $v_B$ is angle-dependent in a lattice model \cite{nahum3}.  $\lt$ may recover continous rotational symmetry at asymptotically low temperatures when one approaches certain quantum critical points, where long wavelength modes (which are weakly affected by lattice anisotropies) dominate: see the discussion in Sec.~\ref{conclusions}.  Higher dimensional random circuits \cite{nahum3} provide lattice models in which it is possible to see the emergence of the higher dimensional minimal surface analytically.

The key difference between higher dimensions and 1D is that curved surfaces appear in the minimization problem for the entanglement of simple compact regions such as a disc of radius $R$ \cite{nahum}. As a result, in order to determine $S(t)$ for times of order $R$ following a quench from an unentangled state, it is necessary to know the full function $\lt(v)$, unlike in 1D where this picture gives a universal piecewise linear scaling form, depending only on $v_E$, for the entanglement of a finite interval.

\section{Spreading operators}

\label{operator_entanglement_definition}

In this section we investigate the production of operator entanglement within the ``footprint'' of a spreading operator.  It had been suggested as an initial toy model of this process that the operator rapidly becomes maximally entangled within the region where it is present \cite{ha}, but we find that this is not the case:  A spreading operator is volume-law entangled, but the entanglement entropy density is well below that of a maximally entangled operator.

To begin with let us recall how entanglement is defined for operators in a spin system with $q$ states per site (e.g. a spin--1/2 system with $q=2$). We may view an operator as a state in a ``doubled'' system with $q^2$ states per site. We may then define the entanglement entropy of the operator using the usual prescription for states \cite{zanardi,prosenpizorn,prosenpizorn2,luitz,dubail}.

The mapping between operators and states can be seen at the level of a single site operator $O_{ab} \ket{a} \bra{b}$, with ${a,b=1,\ldots, q}$. The corresponding state is ${\opket{O} = O_{ab} \ket{a} \otimes \ket{b}}$. The notation $\opket{\ldots}$ indicates that this state lives in the doubled system.\footnote{This mapping between operators and states requires a choice of local basis, but the operator entanglement is independent of this choice.}  

From an operator $A$ on the full system, we obtain a state $\opket{A}$ on $2\times N$ spins, where $N$ is the number of physical spins.  We are free to consider the entanglement of any subset of these $2N$ spins: see Fig.~\ref{fig:Uentanglement}. When it is necessary to distinguish the two sets of $N$ spins we will refer to them as the ``row'' and ``column'' spins (since they correspond to row and column indices of the operator respectively). 

Sometimes it is convenient to group together the row and column spins at a given physical site $i$. For a spin-1/2 chain the four basis ``states'' for an operator at site $i$ may be taken to be the local identity $\opket{\mathbb{I}}_i$ and the three local Pauli operators, $\opket{X}_i$, $\opket{Y}_i$, $\opket{Z}_i$, where we have suppressed normalization constants. In the following we will assume all kets to be normalized.
 
The Heisenberg dynamics of an operator $A(t)$ is
\begin{equation}
    A(t)=U^{\dagger}(t)A(0)U(t) ~,
    \label{eq:operator}
\end{equation}
where $U(t)$ is the time evolution operator. In the ``state'' language this is
\ba
\opket{A(t)}& = (U^\dag\otimes U^T) \opket{A(0)},
\end{align} 
or ${i \partial_t \opket{A(t)}  =\mathcal{H} \opket{A(t)}}$  for Hamiltonian evolution, with ${\mathcal{H} =(\mathbb{I} \otimes H^*}-H\otimes \mathbb{I})$. Since the dynamics occurs separately in each of the two copies, any physical conserved quantity gives rise to two conserved quantities in the operator evolution.

The global identity operator is an eigenstate $\opket{\mathbb{I}}$ of this dynamics with a very simple entanglement pattern: each row spin is maximally entangled with the column spin at the same position $i$. The overlap between a given operator and this state is proportional to $\tr A$ and  is conserved over time. In the following  we consider traceless operators, $\tr A(t) =0$. Operators with a nonzero trace generate entanglement more slowly, due to the overlap with $\opket{\mathbb{I}}$, see Sec.~\ref{tracesection}.

 We will be particularly interested in operators that are initially local \cite{rss}: $A(0)$ acts nontrivially on only one or a few nearby sites of the chain, and is the outer product of a traceless local operator on those sites and the identity at all other sites.  Under the unitary time evolution, this initially local operator spreads and becomes increasingly entangled.  It is the dynamics of this operator spreading and entangling that we explore. 

For  a chaotic Hamiltonian at energies that correspond to high temperature (or more generally chaotic evolution with a Hamiltonian that is not constant in time), generically any
product \textit{state} $\ket{\psi}$
 will become more entangled under the dynamics. For operators, on the other hand, there is always the special product operator $\mathbb{I}$ that is time-independent and does not generate entanglement between different sites. 
(In the presence of conserved quantities there may be other product operators that  commute with $U(t)$ and thus  do not evolve.)
Even when $A(t)$ is traceless,  so that its overlap with $\opket{\mathbb{I}}$ vanishes, $A(t)$ may \textit{locally} consist of a product of identities in some region.
Nontrivial dynamics occurs only in regions where the operator is \textit{not} the local identity:  these ``active'' regions spread ballistically in to the initially ``quiet'' identity regions at the butterfly speed  $v_B$ \cite{rss}.  Operator entanglement is generated only within and at the edges of active regions.
The point that operator entanglement is generated only within the $v_B$ lightcone has also been made recently in Ref.~\cite{xuswingle2018}.

For a given initially-local operator $A(t)$, at time $t$ we can identify the region where it is ``present'' as the locations where a substantial fraction of the operator's total weight consists of local non-identity operators.  
 Quantitatively, we may define the region where the operator is present via an out of time order correlator,\footnote{We may also write $\mathcal{C}_i$ in the form
\be
\mathcal{C}_i(t) = - \f{1}{8 \tr A^\dag A} \sum_{\mu=x,y,z} \tr \, [\sigma^\mu_i, A(t)]^2.
\ee
}
\be
{\mathcal{C}_i(t) = \opbra{A(t)} P_i \opket{A(t)}},
\ee
where $P_i$ is the local projector onto ``states'' \textit{orthogonal} to the identity (locally, ${P_i = 1-\opket{\mathbb{I}}_i \opbra{\mathbb{I}}_i}$).   $\mathcal{C}_i(t)$ is small outside the lightcone defined by $v_B$, and $\mathcal{C}_i(t)$ saturates to an $O(1)$ constant deep inside the lightcone (due to an equilibration of the local ``structure'' of the operator).  For an initially local operator in a thermodynamically large system, the local equilibration of the operator is to  infinite temperature,\footnote{This follows from the fact at almost every site  the initial state $\opket{A(t)}$ resembles the local identity state $\opket{\mathbb{I}}_i$. This means that the densities of conserved quantities are the same as in the infinite temperature state, up to $1/N$ corrections.} which means that for a spin-1/2 system at infinite temperature $\mathcal{C}_i(t)\rightarrow 3/4$ inside the lightcone. $\mathcal{C}_i$ has a front --- or in 1D, two fronts ---  in which it transitions between this value and zero. The width of these fronts is parametrically smaller than the size of the operator when $t$ is large, ensuring that $v_B$ is well-defined. 

For the continuous time evolution  in Eq.~\ref{eq:ham}, we find numerically that the front does broaden, but sublinearly with time. This is consistent with a scaling picture obtained from  discrete time dynamics \cite{nahum3,keyserlingk}, and is confirmation that front broadening also occurs in continuous time models.

\subsection{Entanglement of a spreading operator}
\label{spreadingoperatorsubsection}

\begin{figure}[t]
\centering{
\includegraphics[width=.49\linewidth]{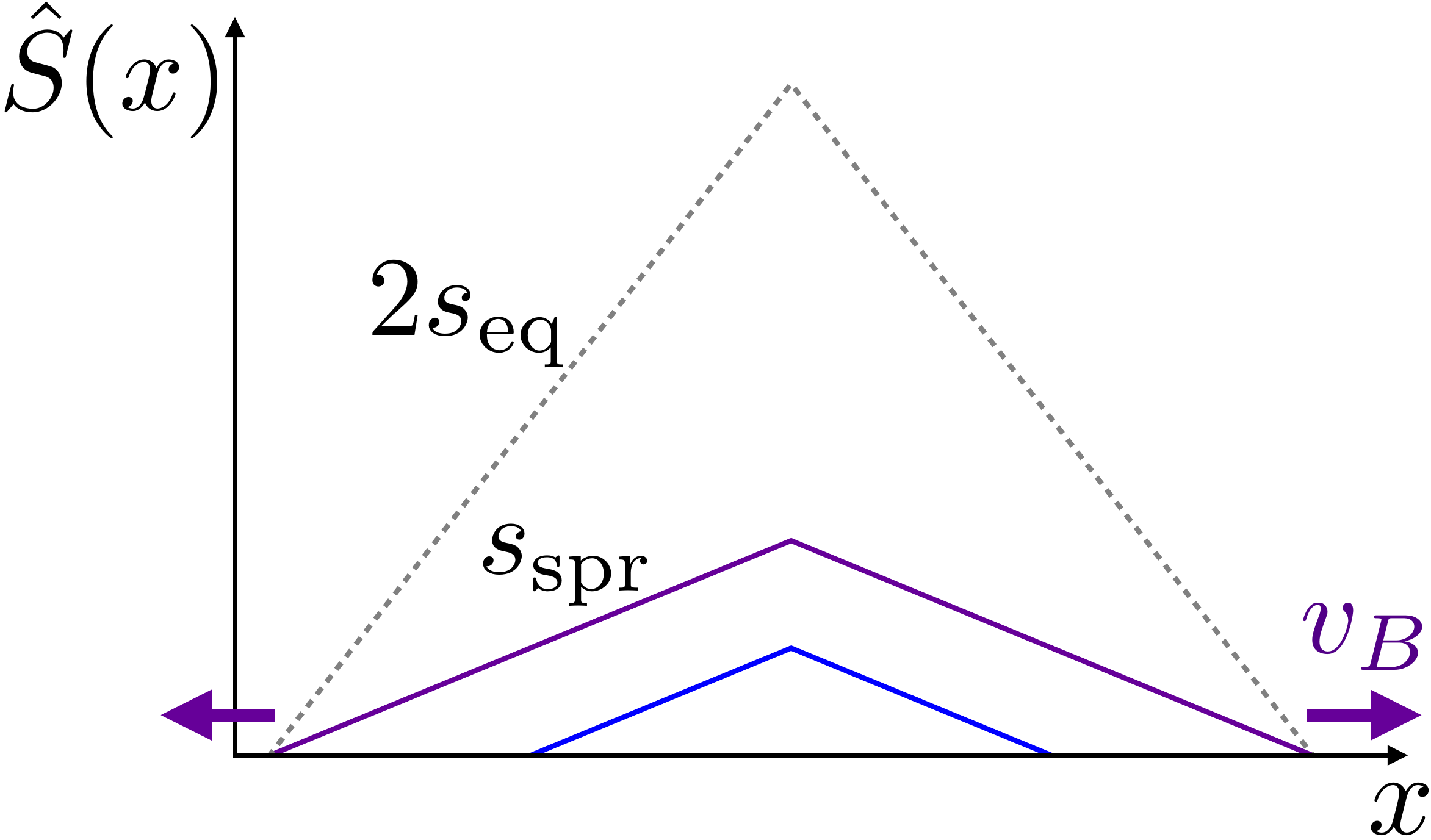}
\includegraphics[width=.49\linewidth]{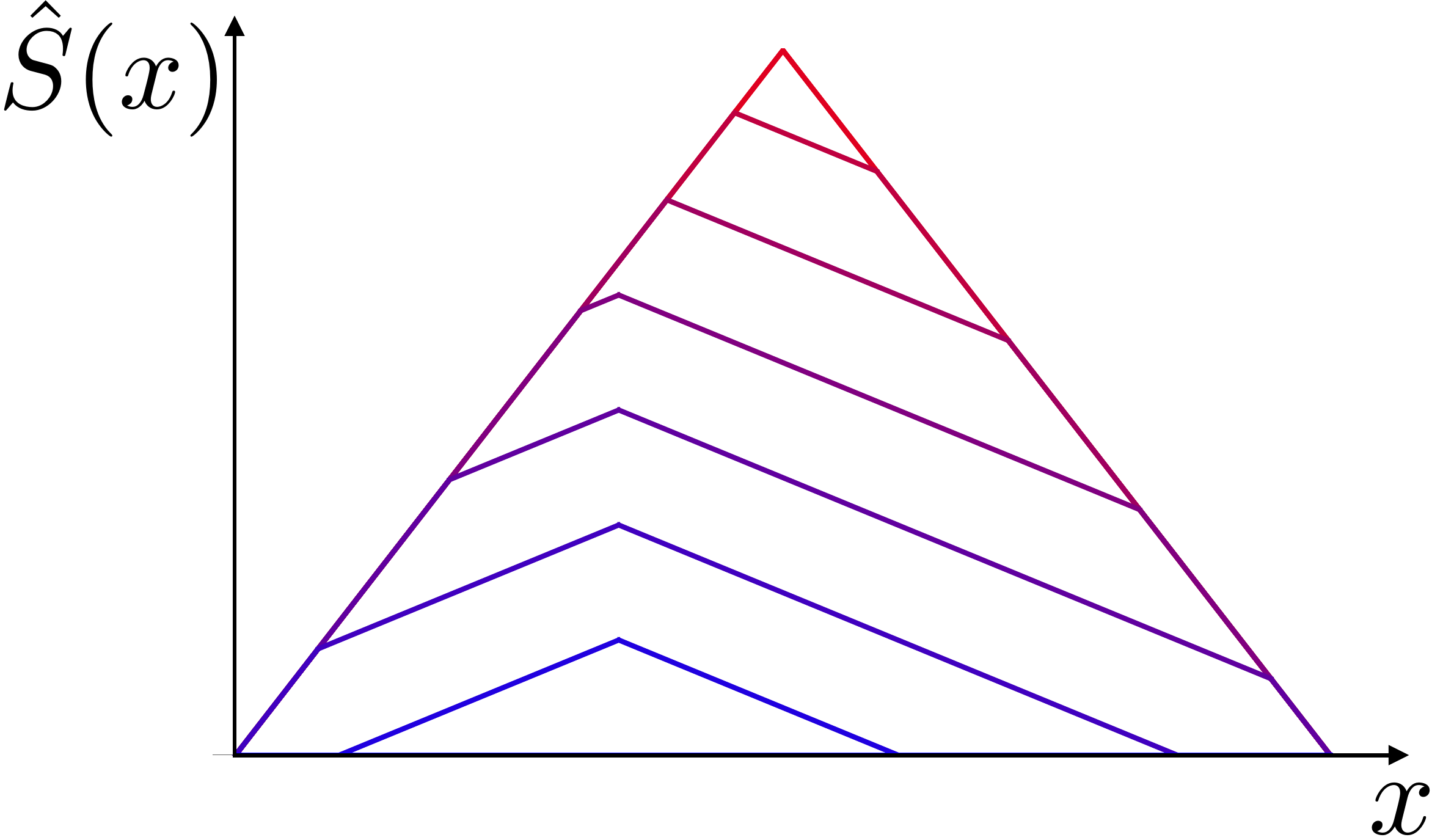}
}
\caption{
Left: cartoon of operator entanglement $\hat S(x)$ for a spreading operator in an infinite chain at two successive times, showing an expanding pyramid with a gradient $\sspr$ that is smaller than the gradient $2\seq$ for a ``typical'' operator with the same spatial footprint (shown dashed).
Right: cartoon of $\hat S(x)$ 
for a spreading operator in a \textit{finite} chain at successive times. Once the operator reaches a system boundary, it becomes maximally entangled close to the boundary, eventually saturating to the pyramid profile with slope $2\seq$.
} 
 \label{fig:S_op_cartoon}
\end{figure}

We propose a coarse-grained phenomenology for entanglement of a spreading operator which is inspired and supported by numerical results for the nonintegrable Ising chain.  
We first discuss the 1D case via a simple generalization of Eq.~ \ref{eq:state} for $\partial S/\partial t$,
where we find an ``expanding pyramid'' form for the entanglement across a spatial cut through the operator (Fig.~\ref{fig:S_op_cartoon}, Left).
After presenting the numerical results we then  describe the equivalent spacetime  picture, which generalizes simply to higher D.

\begin{figure}[b]
\centering{
\includegraphics[width=0.55\linewidth]{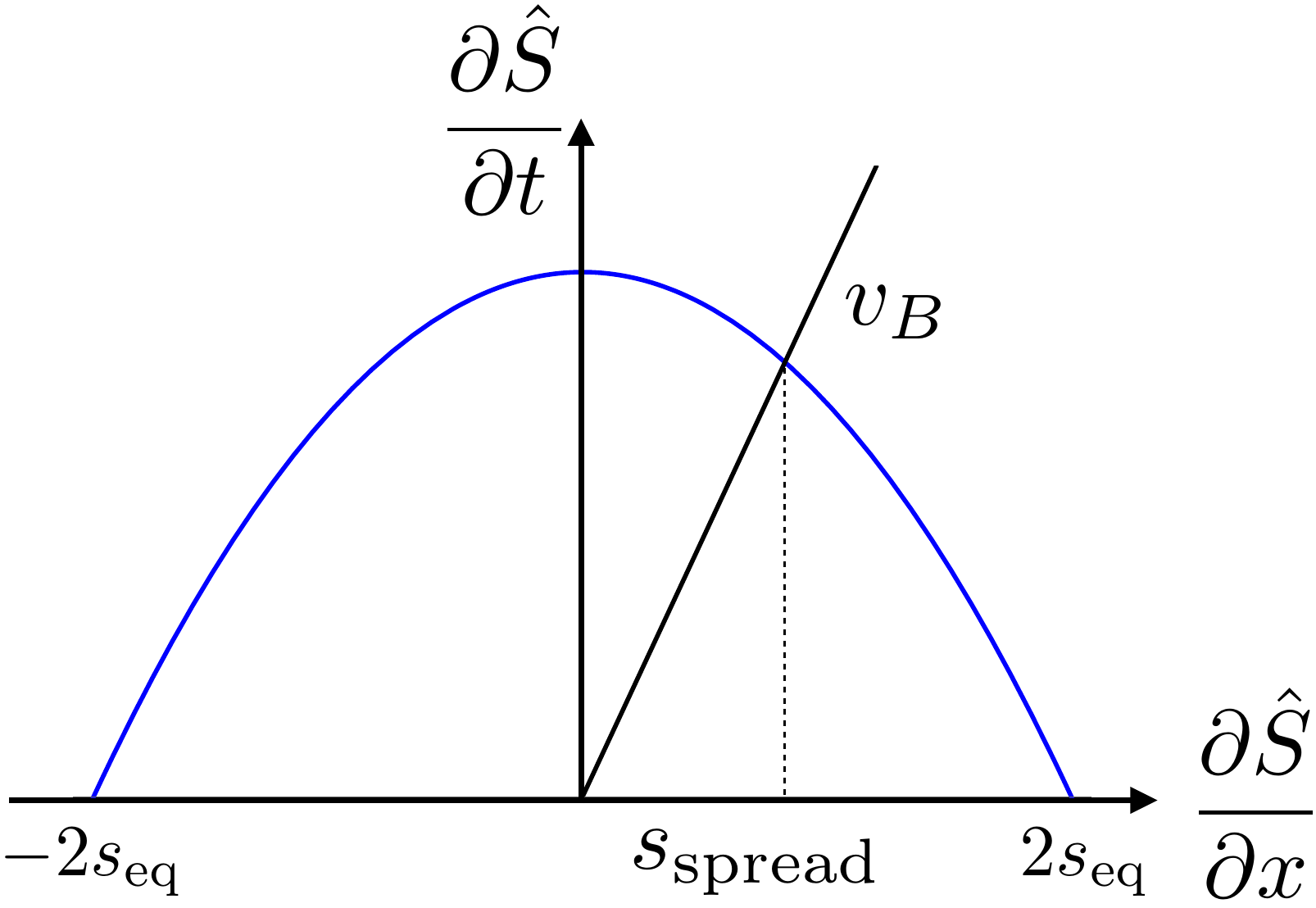}
}
\caption{
Schematic of Eq.~\ref{eq:sspreq}, which gives the slope $\sspr$ of the entanglement $\hat S(x)$ for  a spreading operator in 1D.
Blue curve is ${2\,\Gamma\big( 2^{-1} \partial \hat S/\partial x \big)}$ and straight line is ${v_B  \partial \hat S/\partial x}$.
} 
 \label{fig:ssprfig}
\end{figure}

 The two fronts of the operator  move away from the origin at $v_B$, the butterfly speed. Let  $\hat S(x,t)$ denote the operator entanglement across a cut at spatial position $x$.
This vanishes when $x$ lies outside of the active part of the operator where the operator consists just of local identities.  Within the ``footprint'' of the operator (between the two fronts), we propose that 
\begin{equation}
    \frac{\partial \hat S}{\partial t} = 2\seq~ \Gamma \lf\f{1}{2} \frac{\partial \hat S}{\partial x}\ri,
    \label{eq:operator}
\end{equation}
Here $\Gamma(s)$ is the growth rate defined above for states, and in this section $\seq$ refers to the entropy density for states at \textit{infinite} temperature, since initially local operators equilibrate to infinite temperature.
(Note that the above formula for states, Eq.~\ref{eq:state}, implies Eq.~\ref{eq:operator}  for  operators of the special form $\ket{\psi}\bra{\psi}$ with $\ket{\psi}$ at the appropriate energy density; of course this does not correspond to an initially local operator.)
Below we will discuss the spacetime interpretation of Eq.~\ref{eq:operator}.

The maximal entanglement slope for the operator, for which the right hand side of (\ref{eq:operator}) vanishes, is twice that for the state, $2\times \seq$. For the infinite temperature spin-1/2 chain studied numerically below, this is two bits per site.

\begin{figure}[t]
\centering{
\includegraphics[width=\linewidth]{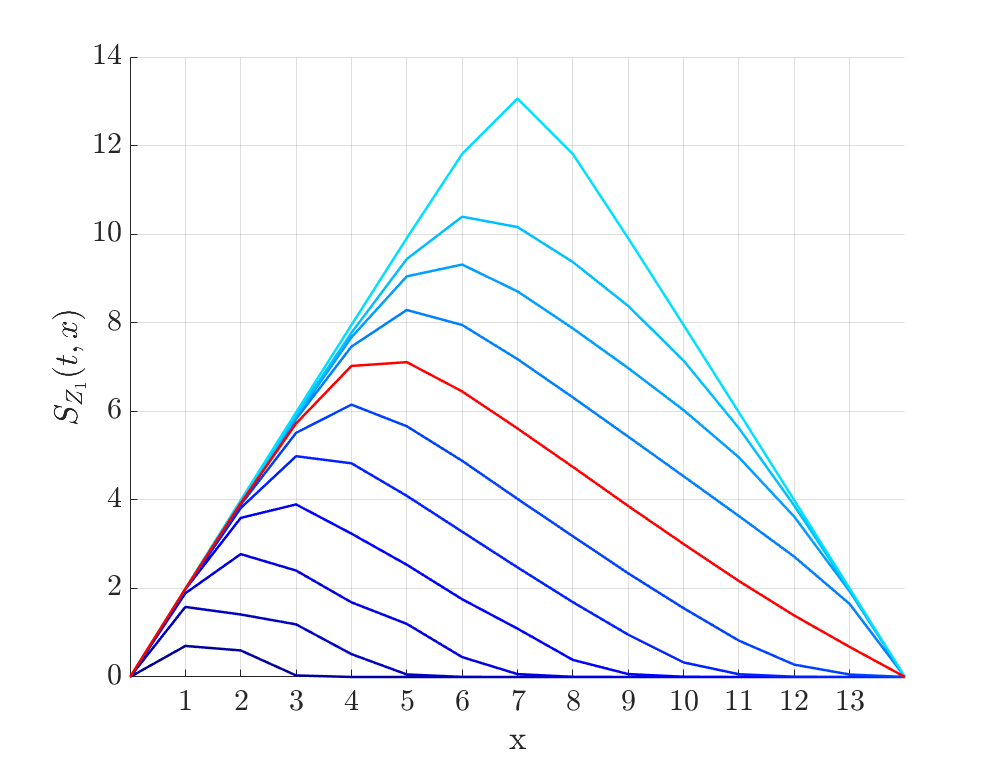}
}
\caption{
Time dependence of the operator entanglement, across a cut at bond $x$, for the time-evolved Pauli matrix $Z_1(t)$ in a chain of length $L=14$. The times shown are $t=1,2,\ldots,10$ and $t=100$. The right-hand section of the $t=7$ data (red line) has a slope close to our  estimate of $\sspr$. The slope at asymptotically late times is close to 2 bits per site, as expected for a thermalized operator.
}  \label{fig:SZ1}
\end{figure}

In the scaling limit, the entanglement profile of an initially local operator (initially located at the origin) will then be the pyramid 
\be\label{operator_pyramid}
\hat S(x,t)=\sspr (v_B t - |x|),
\ee
for $|x|\leq v_B t$, where  where $\sspr<2$ is the solution to the equation\footnote{Since $\Gamma''\leq0$ there is only one such solution, and this entanglement profile is  dynamically stable.}
\be\label{eq:sspreq}
v_B \, \sspr= 2 \, \Gamma \lf \f{\sspr}{2} \ri.
\ee
This is illustrated in Fig.~\ref{fig:ssprfig}.  
The entanglement profile $\hat S(x,t)$ of the spreading operator consists of four linear sections: the two regions outside of the operator's edges to the left and right where $\hat S=0$, and the two linear sections on either side within the spreading operator, as illustrated in Fig.~\ref{fig:S_op_cartoon} (Left).  At the points where these linear sections meet, the exact entanglement profile is smoothly rounded out due to higher order terms in the gradient \cite{nahum} that are ignored in this leading coarse-grained entanglement dynamics (3), as we will see in the numerics below.

In this scenario the entanglement gradient $\sspr$ of the spreading operator is necessarily less than that of a maximally entangled operator whenever $v_B>0$.  When an edge of the operator reaches the end of a finite spin chain, then the operator stops spreading, 
allowing the region of the operator adjacent to this end to become maximally entangled, as illustrated in Fig.~\ref{fig:S_op_cartoon} (Right).

We have  explored the entanglement of spreading operators numerically in the quantum chaotic Ising spin chain with longitudinal and transverse fields, Eq.~(\ref{eq:ham}), with $L$ sites for $L$ up to 14. We  diagonalize this Hamiltonian exactly to obtain the operator dynamics.  We will present results for the spreading of the initially local operators $Z_1$ and $Z_{L/2}$, one of which starts near the end of the chain and the other near the center. Other initially local operators behave essentially the same as these two examples.

The behavior of $\hat S(x,t)$ for the operator $Z_1(t)$ in a chain of length $L=14$ is shown in  Fig.~\ref{fig:SZ1}. (The figure shows $t=1,2,\ldots, 10$ and $t=100$.) By starting the operator at the end of the chain, we are able to watch it spread in one direction over a distance of $(L-1)$ sites.  The operator spreads across the chain, while promptly getting locally maximally entangled near the end of the chain where it started.  This sets up the spreading profile with a roughly linear $\hat S$ vs. $x$ over the central region of the chain, and a steady production of entanglement. 

\begin{figure}[t]
\centering{
\includegraphics[width=\linewidth]{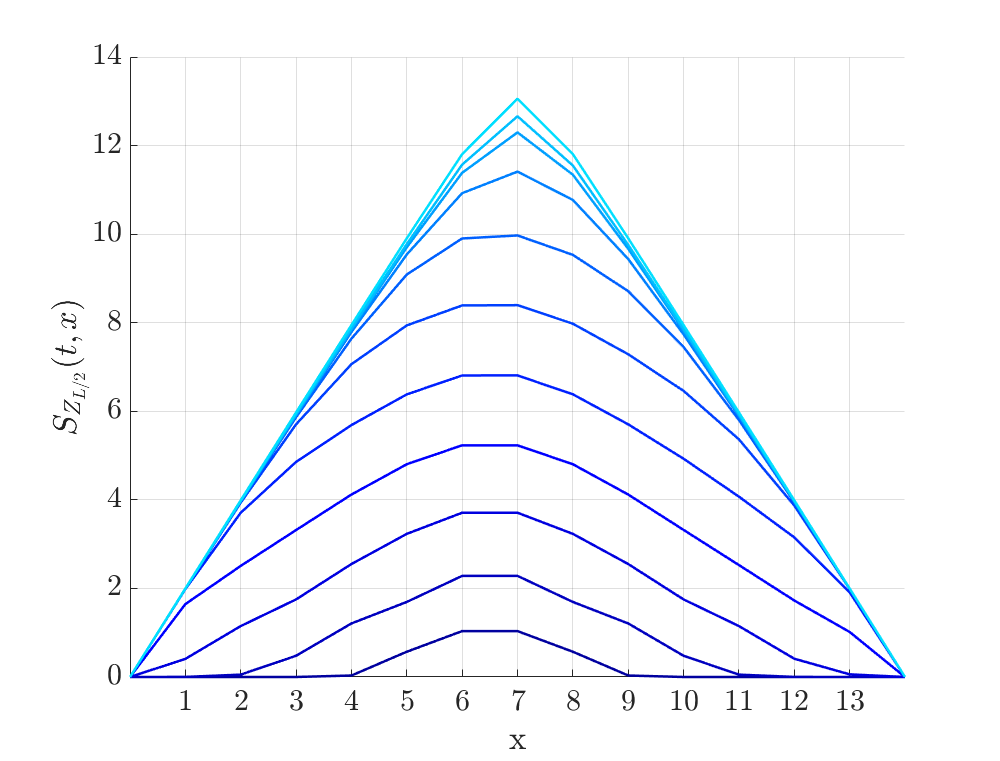}
}
\caption{
Operator entanglement of the operator $Z_{L/2}(t)$ which starts near the centre of the chain, for times ${t=1,2,\ldots,10}$ and $t=100$.} 
 \label{fig:SZmid}
\end{figure}

We obtain the steady-state slope $\sspr = |\partial \hat S/\partial x|$ by measuring the slope of the right-hand linear section of the entanglement profile. The $L=14$ data at $t=L/2$ shows a slope $\sspr \sim 0.87$. Appendix~\ref{app:extrapolationsspr} compares results for $L=8,10,12,14$ to give an idea of the finite-size effects. We estimate that $\sspr$ lies in the range 
\be 
\sspr \in (0.9,1.0).
\ee
This is well below the maximal entanglement of \textit{two} bits per site, which is attained in the long time limit (after reaching both ends of the chain).

The entanglement of $Z_{L/2}(t)$ is shown in Fig.~\ref{fig:SZmid}. The features are similar but finite size effects set in sooner.

\subsection{Spacetime picture}

\begin{figure}[t]
\centering{
\includegraphics[width=0.5\linewidth]{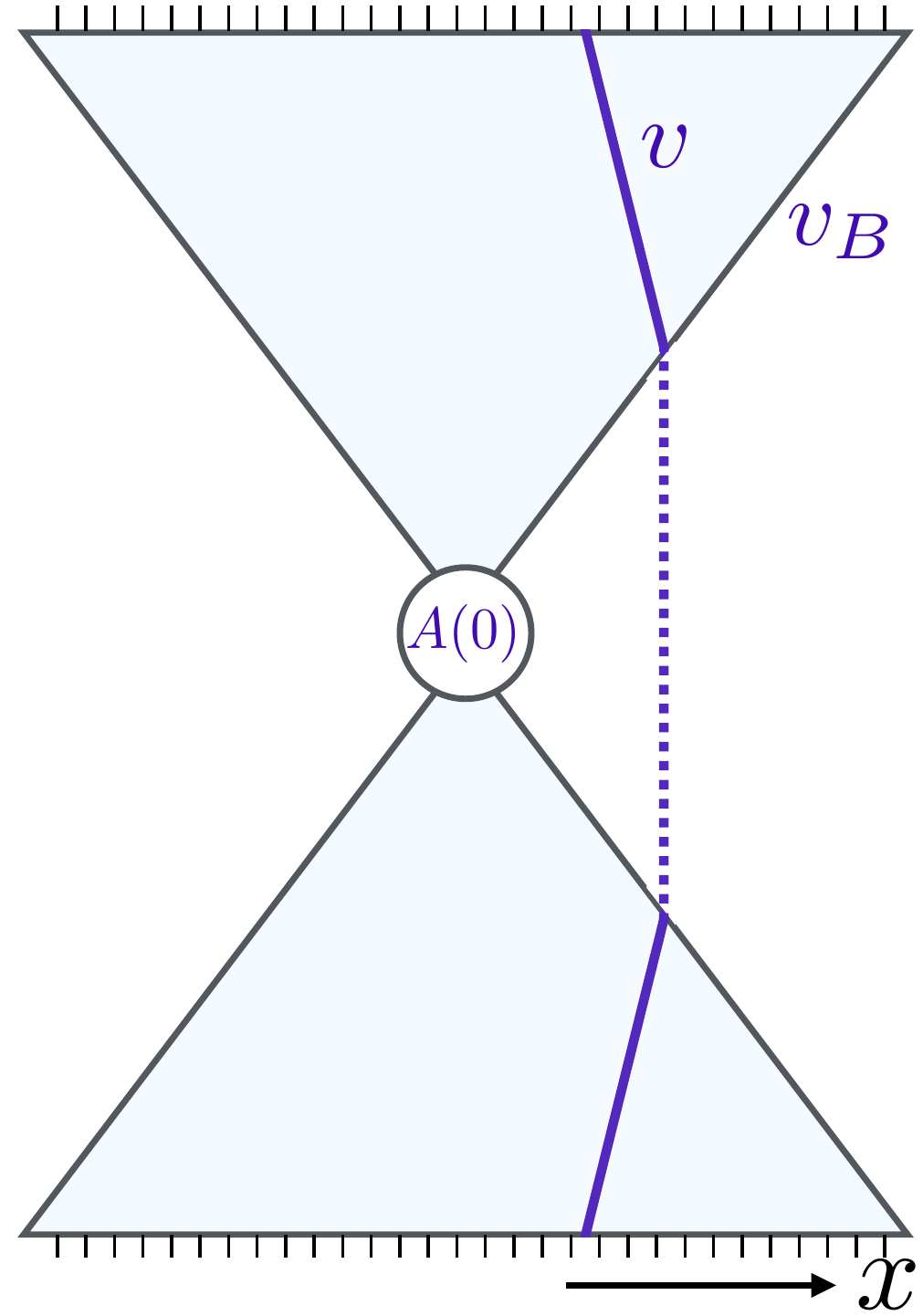}
}
\caption{
Minimization for spreading operator giving Eq.~\ref{spreadingoperatorminimization}.}  \label{fig:spreadingoperatorcut}
\end{figure}

Let us consider $\hat S(x,t)$ for a spreading, initially localized operator in the minimal surface picture. We assume that the same effectively local description, in terms of a line or surface tension $\lt(v)$, applies within the doubled spacetime patch representing the action of both $U(t)^\dag$ and $U(t)$ in the Heisenberg evolution of the operator. We  further assume that the effects of operator spreading can be taken into account simply by ``truncating'' the circuit to the lightcone defined by $v_B$ (see Fig.~\ref{fig:spreadingoperatorcut}). That is, we assume that unitaries in $U(t)$ outside this lightcone effectively cancel with their partner in  $U(t)^\dag$ to leave the identity.

By symmetry, the minimal cut configuration determining $\hat S(x,t)$ for $x>0$ is then that shown in  Fig.~\ref{fig:spreadingoperatorcut}. This gives
Eq.~\ref{operator_pyramid}, with 
\be\label{spreadingoperatorminimization}
\sspr = 2 \seq \min_v \f{\lt(v)}{v+v_B},
\ee
where $v$ is the inverse slope of the non-vertical sections.
This is  equivalent by (\ref{legendre_transform}) to the expression above in terms of $\Gamma(s)$.

For an operator initiated at the origin we may consider the entanglement of the set of sites within a distance $r$ from the origin.  At short times this is just twice the value above (since we have two separate cuts of the type discussed above) while at late times the entanglement saturates to {\blue $4\seq r$}. This implies that this set of $2r$ sites is fully entangled with the exterior if ${r<v_\text{core} t}$, where {\blue $v_\text{core} = v_B \sspr/(2 s_\text{eq}+\sspr)$}. The speed $v_\text{core}$ sets the size of the ``fully-entangled'' core of the operator. 

The entanglement $S(0,t)$ across the midpoint of an operator started at 0 in an infinite system is ${S(0,t)=\sspr v_B t}$. 
Therefore storing $A(t)$ in matrix product operator form is less expensive than storing a ``thermalized'' operator of the same size, which has entanglement  $2 \seq v_B t$ across its midpoint. 

However $\sspr v_B t$ is still always larger than the \textit{state} entanglement generated by a quench from a product state in the same amount of time, $\seq v_E t$ (see below). Therefore to compute the expectation value $\<A(t)\>$ numerically following a quench from a product state it is likely to be more efficient to use the Schrodinger picture than the Heisenberg picture (time evolving the state rather than the operator). 

This should be contrasted with certain integrable chains (including Ising and XY and perhaps XXZ), in which  the entanglement of a spreading operator grows only logarithmically with time \cite{prosenpizorn, prosenpizorn2,dubail}. For those systems, storing the evolving operator in matrix product form is much more efficient than storing the evolving state.

In general Eq.~\ref{eq:vbconstraint} gives the bounds $\sspr /  \seq \leq \min \{ 2 v_E/v_B, 1\}$ and $\sspr/ \seq \geq 2 v_E / (v_E + v_B)$. For the random circuit in Eq.~\ref{randomstructureexample} these bounds read $2/3 \leq \sspr /  \seq \leq 1$, and the actual value  is $\sspr/\seq  = 2 (\sqrt 2 - 1)\simeq 0.83$,  smaller than but similar to the value  we find numerically for the nonintegrable Ising model.

The picture above generalizes directly to higher dimensions. For an initially local operator in a  rotationally invariant system we must solve a membrane minimization problem in a cone-shaped spacetime region.

\section{Extensions}

\subsection{Higher-gradient corrections}
\label{higher_gradient}

So far we have discussed the leading order coarse-grained entanglement dynamics, 
but subleading effects are needed to understand the more detailed features of our numerics.  It is natural to expect that in many situations the dominant such effects will be described by higher spatial derivative corrections to Eq.~\ref{eq:state} and the comparable formulas for operators.  Explicit calculations  for the higher Renyi entropies, and for the von Neumann entropy in certain limits, show that such higher-derivative corrections are present for random unitary circuits (where the presence of randomness also leads to universal subleading fluctuations).\cite{nahum,zhounahum} 
The first such subleading term is $\partial^2 S/\partial x^2$ with a coefficient that depends on $\partial S/\partial x$ (which is not necessarily small in this regime).  In this section we argue that such higher-derivative corrections explain differences in finite-time entropy growth rates for the various initial states/operators we have considered.

To begin with let us  compare the state entanglement for two different initial conditions.  First, the entanglement following a quench from an initial unentangled product state, $\Psi = \otimes_i \Psi_i$, which we denote  $S_{\otimes \Psi}(x,t)$.   The leading order dynamics is Eq.~\ref{eq:state} with the initial condition ${S_{\otimes \Psi}(x,0) = 0}$.  Second, the state entanglement following a quench from an initial state $\Psi= \Psi_A\otimes \Psi_B$ which is the product of independent Page--random states in the left and right halves of the system.  The initial ${S_{\Psi_A\otimes\Psi_B}(x,0)}$ now has a two-pyramid structure.  For $t< L/(4 v_B)$,  ${S_{\Psi_A\otimes\Psi_B}(x,0)}$ resembles Fig.~\ref{fig:joiningprocess}.

In both cases $\partial S/\partial x|_{x=L/2}$ vanishes for $t>0$, so in the leading order treatment the entanglement across the central bond grows at the same rate,
\ba
S_{\otimes \Psi}(0,t) & \sim 
S_{\Psi_A\otimes\Psi_B}(0,t)  \sim \seq v_E t.
\end{align}
However while the product state initial condition gives a flat entanglement profile, the two-pyramid initial condition gives a positive curvature $\partial^2S_{\Psi_A\otimes\Psi_B}/\partial x^2$ at the central bond, which (from the leading order dynamics) decreases like $1/t$. Therefore if the first subleading correction to $\partial S/\partial t$ is $\partial^2S/\partial x^2$ with a positive coefficient we  expect $\partial S_{\Psi_A\otimes\Psi_B}/\partial t$ to tend to $\seq v_E$ from above like $1/t$.
The available system sizes do not allow us to check this power law, but we do find that $\partial S_{\Psi_A\otimes\Psi_B}/\partial t$ is greater than $\partial S_{\otimes \Psi}/\partial t$ at early times. The time derivatives are shown in Fig.~\ref{fig:joiningprocess}, along with the lattice curvature
$(\partial^2 S/ \partial x^2)_\text{lattice}$ at $x=L/2$.\footnote{ $(\partial^2 S / \partial x^2)_\text{lattice}\equiv {\widetilde S_U(L/2+1) - 2 \widetilde S_U(L/2) + \widetilde S_U(L/2-1)}$.}

\begin{figure}[t]
\centering{
\includegraphics[width=\linewidth]{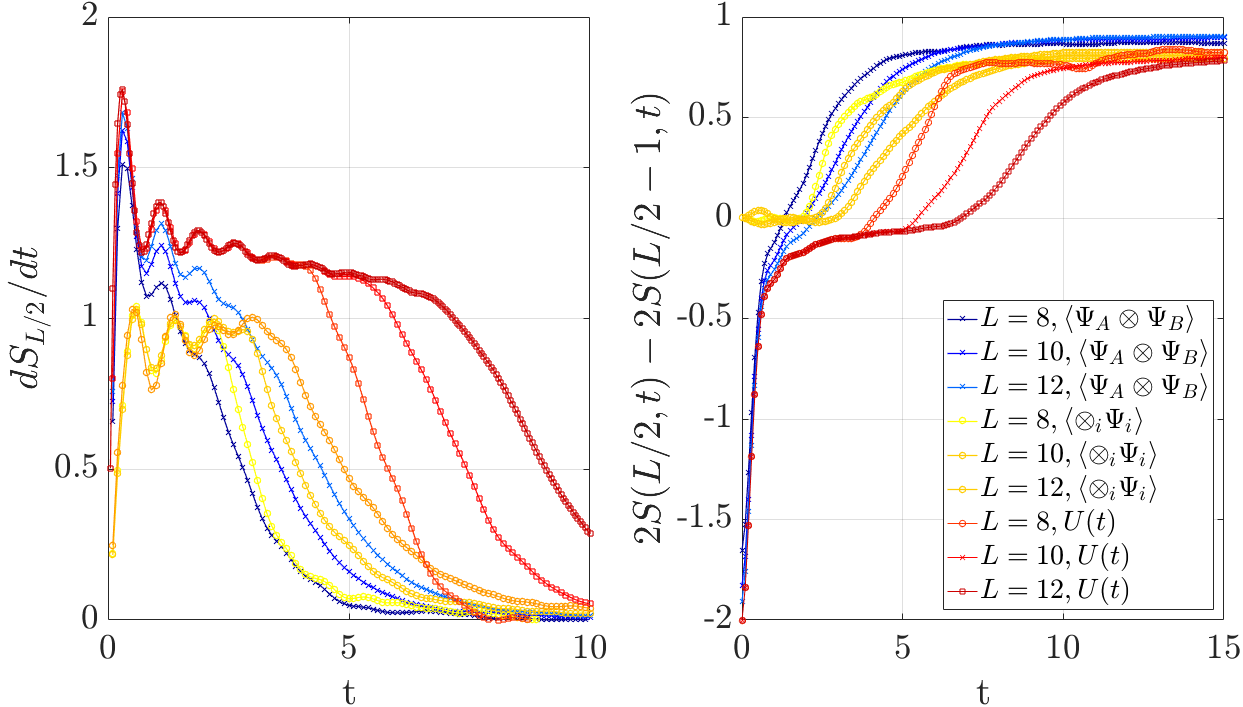}
}
\caption{
Growth rate $\partial S/\partial t$ of entanglement across the central bond (left), 
and  lattice approximation to $\partial^2S/\partial x^2$ (right), for three protocols discussed in Sec.~\ref{higher_gradient}: an initial state that is a product of  two Page-random states in the two halves of the chain, $S_{\Psi_A\otimes\Psi_B}$; an initial state that is an unentangled product state, $S_{\otimes \Psi}$; and the unitary entanglement $S_U(x,L/2,t)$. The positive $\partial^2S/\partial x^2$ at early times, for the first and third of these, is associated with an increased $\partial S/\partial t$.
}  \label{fig:dSdt}
\end{figure}

\begin{figure}[t]
\centering{
\includegraphics[width=\linewidth]{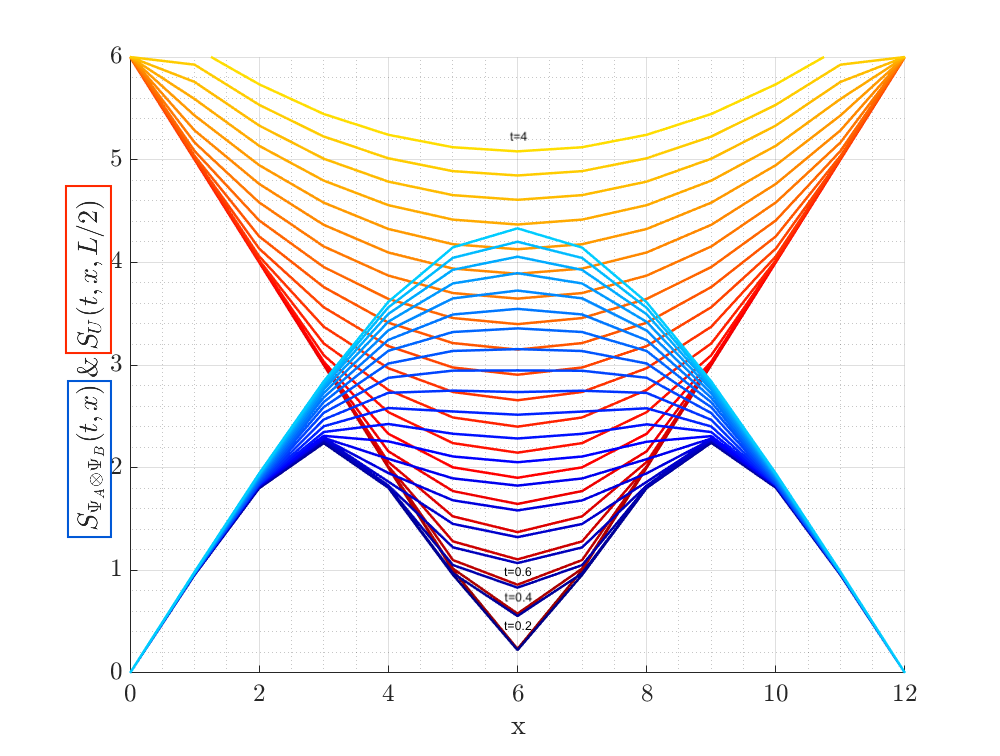}
}
\caption{
Comparison of entanglement growth for (1) a state which starts as a product of two Page-random states, one in each half of the chain, denoted $S_{\Psi_A\otimes\Psi_B}$; (2) the entanglement $S_U(x,L/2,t)$ of the time evolution operator. Times shown are at intervals of 0.2 from $t=0.2$ up to $t=4$.
}  \label{fig:compareunitarystate}
\end{figure}

Interestingly, some support for the idea that the positive curvature in $\partial^2 S/\partial x^2$ is responsible for the increased $\partial S/\partial t$ at early times comes from comparing $S_{\Psi_A\otimes\Psi_B}$ with the entanglement $S_U$ of the time evolution operator (defined in Sec.~\ref{scaling_picture_section}). 
Let us define ${\widetilde S_U(x,t) \equiv S_U(x,L/2,t)}$.  At leading order $\widetilde S$ obeys the same equation as the state entanglement, ${\partial_t \widetilde S_U(x,t) = \seq \Gamma\lf \partial_x \widetilde S(x,t) \ri}$,
and the spacetime picture suggests that the subleading corrections will also be the same. The initial condition is $\widetilde S_U(x,0) = \seq |x-L/2|$. Therefore, within the central region of the chain (for ${|x-L/2|< L/4}$), the initial entanglement $\widetilde S_U(x,0)$ is identical, in the scaling limit, to the state entanglement $S_{\Psi_A\otimes\Psi_B}(x,0)$.  Therefore a first check that these subleading corrections to the hydrodynamics make sense is that  (for  $t\lesssim \f{L}{4 v_B}$) the growth rates should be close for the two quantities.
Fig.~\ref{fig:compareunitarystate} compares the entanglement growth for the state and the unitary, showing approximate agreement at early times. This is clearer in Fig.~\ref{fig:dSdt} (left), where three system sizes are shown. Note that finite time effects set in later for $S_U$ because the saturation value of the entanglement is larger.

The time-dependent growth rates in Fig.~\ref{fig:dSdt} (left) are also consistent with all three growth rates tending to the same constant $v_E$, which is required by the leading order dynamics. We estimate $v_E$ to be in the range $(0.95,1.1)$, as quoted in Sec.~\ref{scaling_picture_section}.  This is consistent with a previous estimate for the same model~\cite{ms}.

Similar corrections to those discussed above arise for the entanglement of operators evolving in the Heisenberg picture. 
Consider an operator $A(t)$ which starts out as a random product of Pauli matrices $B_i$ at the sites, $A(0) = \prod_i B_i$. At leading order ${\f{\partial \hat S_A (x,y, t)}{\partial t}  = \seq \Gamma \lf  \f{\partial \hat S_A }{\partial x}    \ri
+ \seq \Gamma \lf  \f{\partial \hat S_A }{\partial y}    \ri}$,
with the initial condition $S_A(x,y,0) = \seq |x-y|$. For $t> 0$ this nonanalyticity is rounded out, giving positive curvatures $\partial^2 \hat S_A/\partial x^2$ and $\partial^2 \hat S_A/\partial y^2$ which again decay in time like $1/t$. This suggests that $\partial S_A(x,x,t)/\partial t$ converges to $2 v_E \seq t$ from above, with an $O(1/t)$ correction.

\subsection{Traceful operators}
\label{tracesection}

The trace of $A$ determines the overlap, conserved in time, between $\opket{A}$ and an unentangled eigenstate of the dynamics, namely $\opket{\mathbb{I}}$. If the squared overlap is $p$, a plausible cartoon for the reduced density matrix spectrum of a spreading operator is $\{p \} \cup \{ (1-p) \lambda_i \}$, where $\{\lambda_i\}$ is the spectrum for the traceless part of the operator.
We do not expect this to be exact (it would be exact if all the Schmidt states of the traceless part of $A(t)$ corresponded to traceless operators)  but it may capture the leading scaling of the Renyi entropies. According to this ansatz, at long times the von Neumann entropy of $A(t)$ is reduced from the scaling form above for a traceless operator (Sec.~\ref{spreadingoperatorsubsection})  by the factor $(1-p)$, while the higher Renyi entropies $S_n$ \textit{saturate} to the order one constants $n (n-1)^{-1} \ln 1/p$ at late times.\footnote{This picture suggests that at the front of the operator, where $\mathcal{C}_i(t)$ becomes small, the von Neumann entropy growth rate may tend to zero in the same manner as  $\mathcal{C}_i(t)$.}  This is an example of the higher Renyi entropies behaving very differently from the von Neumann entropy.

An operator with nonzero trace is a ``cat''-like superposition of two pieces with very different dynamics. Similar phenomena arise for states.  For example the state $\ket{\Psi} =\alpha  \ket{\Psi_1} + \beta \ket{\Psi_2}$, where $\ket{\Psi_1}$ and $\ket{\Psi_2}$ have macroscopically different entanglement profiles, is a non-generic state which will not obey Eq.~\ref{eq:state} in general. Instead the above ansatz suggests that we should compute the entanglement dynamics of $\ket{\Psi_1}$ and $\ket{\Psi_2}$ separately and take the appropriate weighted average.

\begin{figure}[t]
\centering{
\includegraphics[width=0.9\linewidth]{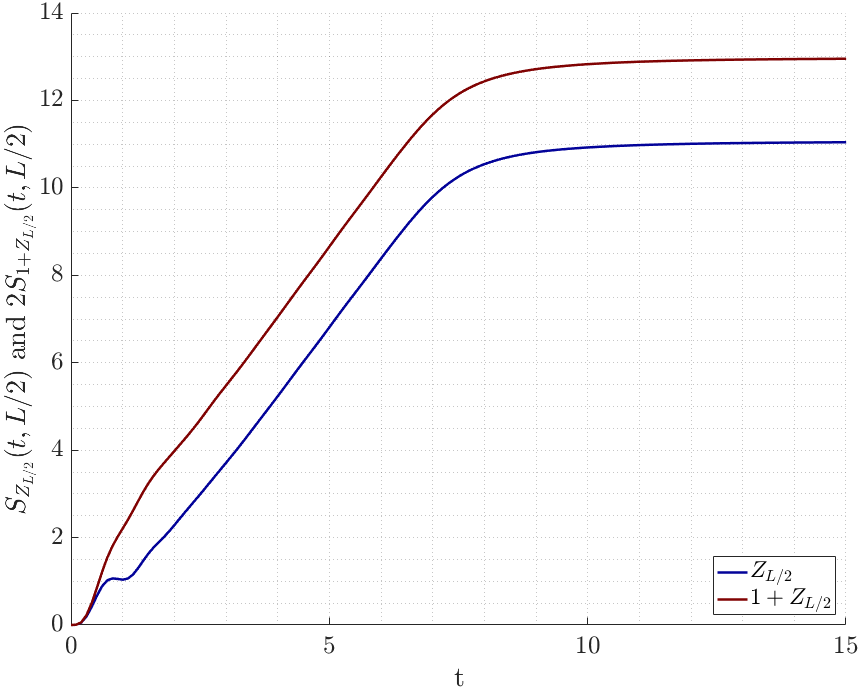}
}
\caption{
Comparison of operator entanglement (across  central bond of  chain) for traceless operator $Z_{L/2}(t)$ and traceful operator $1+Z_{L/2}(t)$. The entanglement of the latter has been multiplied by two.
}  \label{fig:tracefulop}
\end{figure}

In  Fig.~\ref{fig:tracefulop} we show the time dependence of the entanglement for two different initial operators, $Z_{L/2}$ and ${\mathbb{I} + Z_{L/2}}$, across the central bond of the chain. $\hat S$ has been multiplied by two for the second operator, since according to the above ansatz $\partial \hat S/\partial t$ for $Z_{L/2}$ should be twice that of ${\mathbb{I}+ Z_{L/2}}$ when $t\gg 1$. There are transient effects at short time, and finite $L$ effects at late time, but the slopes of the two plots are similar at intermediate times, consistent with the above hypothesis.

\section{Outlook}
\label{conclusions}

We have discussed a simple theory for entanglement generation in nonintegrable many-body systems at leading order in time.
We have given evidence that the minimal path and its line tension $\lt(v)$ are well defined objects in realistic models (not only in analytically tractable models built from random unitaries), and shown that they are related in a simple way to an entanglement generation rate $\Gamma(s)$.
We have shown that this picture unites the entanglement properties of states and operators. 

In general $\lt(v)$ depends on the specific dynamics. However, this function can become universal at low temperatures. Consider for example a quantum critical point (above 1D, to avoid special features of 1D CFTs). Schematically, if we tune a given microscopic model to this quantum critical point, then at low energies it is described by the universal fixed point Lagrangian associated with the quantum critical point, perturbed in general by an infinite number of irrelevant scaling fields. The values of these irrelevant couplings depend on microscopic details, and at high temperature they will have a strong effect on $\lt(v)$. However at low temperature we expect to be able to integrate out modes at frequencies larger than $T$. In this process the leading irrelevant coupling is renormalized to a small value of order $T^{|y_\text{irr}|/z}$, where $y_\text{irr} <0$ is the leading irrelevant exponent and $z$ is the dynamical exponent. Since the Lagrangian is approximately universal in this limit,  it is natural to expect that the function $\lt$ also approaches a universal form, characteristic of a specific universality class of quantum critical point, with a leading correction of order $T^{|y_\text{irr}|/z}$. Is it possible to obtain $\lt$ for paradigmatic quantum critical points such as Ising? 

It would also be interesting to investigate the low temperature entanglement dynamics of theories which flow to a free fixed point, with RG--irrelevant interactions.
 The  interactions are presumably ``dangerously irrelevant'' as 
they are necessary to break the special structure of the free theory (for which a quasiparticle picture is expected) and to establish a well-defined $\lt(v)$ beyond some scale that diverges when $T\rightarrow 0$.

In the future, more precise numerical determinations of $\lt$ for various models at finite temperature should also be possible,  taking subleading effects into account and using larger system sizes. 
 It is also interesting to ask whether  any $\lt(v)$ that satisfies the constraints laid out in Sec.~\ref{scaling_picture_section} can be obtained for some Hamiltonian, or whether there are further restrictions.

It would be  useful to test the scaling picture and the constraints proposed in Sec.~\ref{scaling_picture_section} in a wider range of settings.
It would also be useful to have further analytical results for the von Neumann entropy. At present, detailed results for the von Neumann entropy --- as opposed to the higher Renyi entropies, which are more analytically tractable --- are possible only in limits such as for large local Hilbert space dimension, or alternatively for dynamics of a restricted type.

There are also basic conceptual questions to answer.
We have proposed that the scaling picture applies for ``generic'' initial states with a given entanglement profile. However we have not attempted to make precise what ``generic'' means in this context, and this is an important task for the future.   
In addition, we have not discussed here the effect of conservation laws (which may be incorporated into solvable models \cite{conservationlaws1,conservationlaws2}) on entanglement growth.
A  precise characterization of subleading corrections to the dynamics would also be useful even in the absence of conservation laws. Among other things, this may reveal subtle differences between the dynamics of $S_\mathrm{vN}$ and the higher Renyi entropies that are not captured in our leading order treatment.

\begin{acknowledgments}
AN thanks Jeongwan Haah, Jonathan Ruhman, Sagar Vijay and Tianci Zhou for collaboration on related projects and very valuable discussions.
We also thank John Chalker, Fabian Essler, Vedika Khemani, Mark Mezei and Douglas Stanford for very useful discussions.
AN acknowledges EPSRC Grant No.~EP/N028678/1. 

\end{acknowledgments}

\appendix

\section{Additional data for $S_U$}
\label{app:SUdata}

\begin{figure}[t]
\centering{
\includegraphics[width=0.95\linewidth]{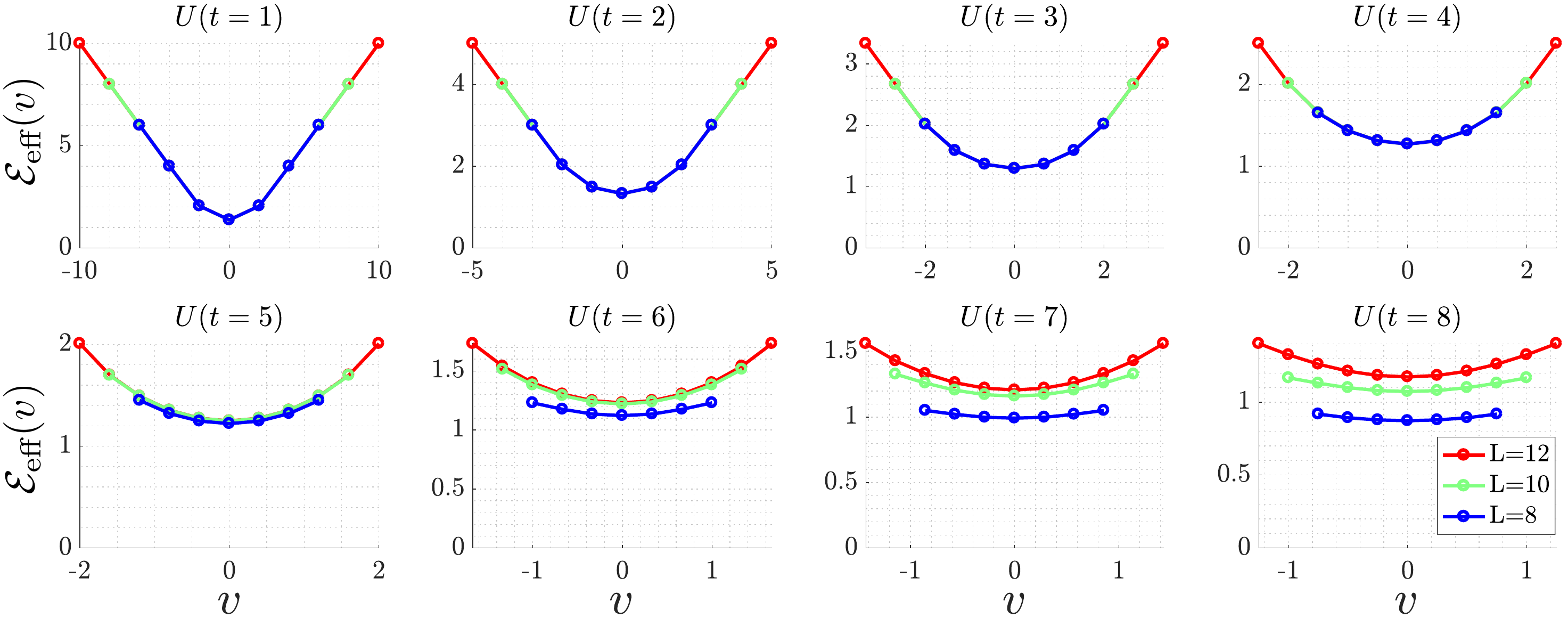}\\ \vspace{4mm}
\includegraphics[width=0.95\linewidth]{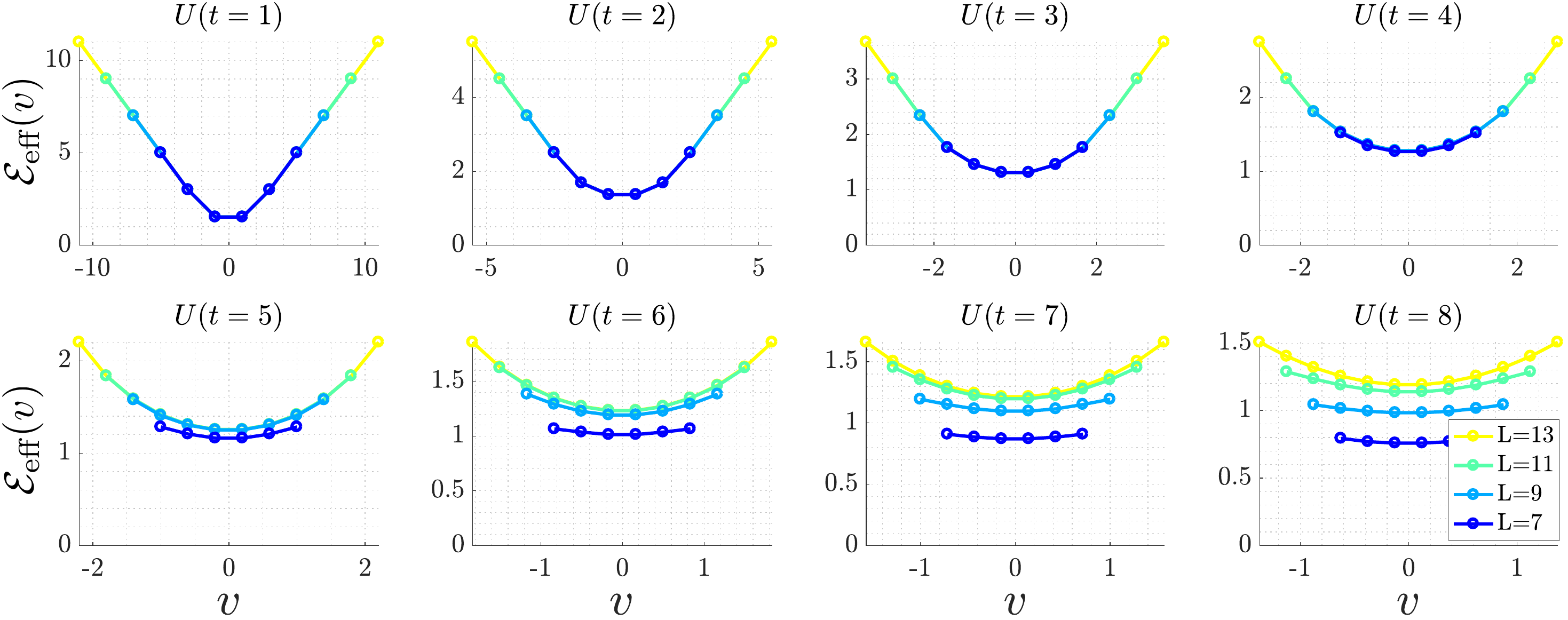}
}
\caption{Top: $\lt_\text{eff}(v)$ (Eq.~\ref{eeffdefn}) for various times in a system of size $L=12$. We take $(x+y)/2=L/2$. Note the scales differ. Bottom: same for $L=13$.}  
 \label{evpanel}
\end{figure}

In Figs.~\ref{evpanel} we show data for  $\lt_\text{eff}(v)$ (obtained from the operator entanglement of $U(t)$, Eq.~\ref{eeffdefn}) for $t=1,\ldots, 8$ for $L=12$ and $L=13$, showing the onset of finite size effects when $t$ becomes large at fixed $L$.  For large enough $t$ the minimal path travels to the spatial boundary of the system and has energy $\seq L$ (to leading order in $L$),  so that the finite-size estimate $\lt_\text{eff}(v)$ tends to zero at late time like $L/t$.
 To minimize finite size effects, in Sec.~\ref{scaling_picture_section} we showed data up to $t=6$ ($L=12$) and $t=7$ ($L=13$). 
 
\begin{figure}[t]
\centering{
\includegraphics[width=0.95\linewidth]{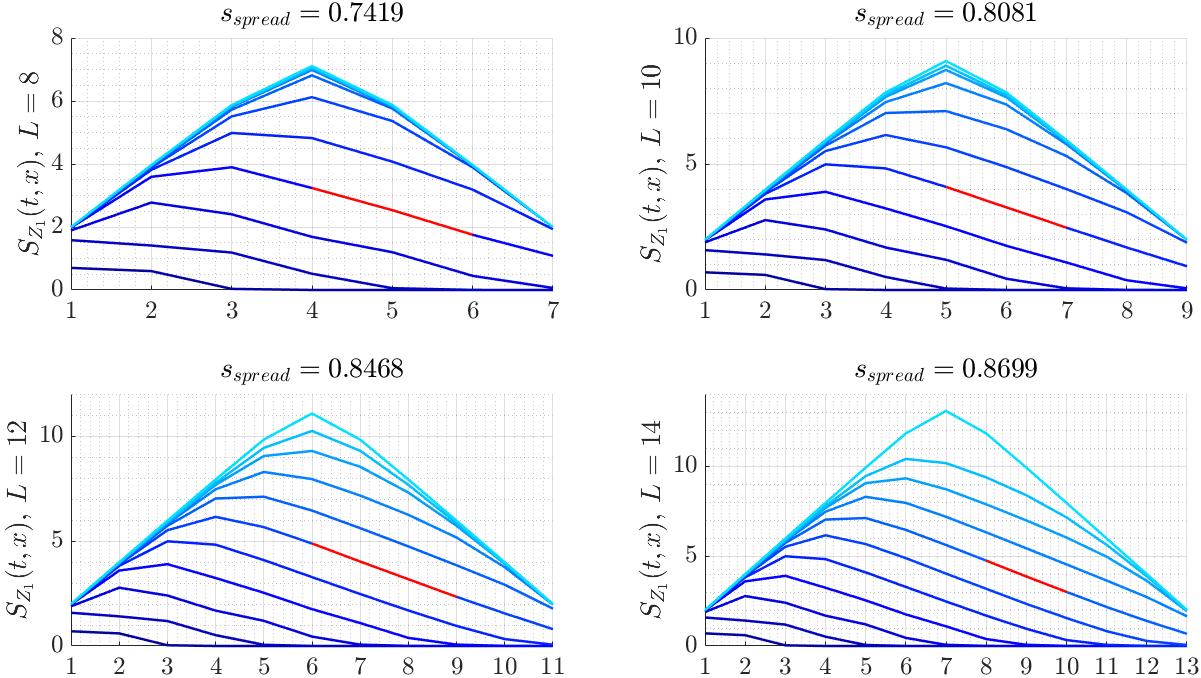}
}
\caption{
Operator entanglement entropy $S_{Z_1}(x,t)$ for the time-evolved Pauli matrix $Z_1(t)$ in systems of size ${L=8,10,12,14}$ for times $t=1,2,\ldots, 10$ and $t=100$. Slopes $-\sspr$ are estimated at $t=L/2$ for the red sections.}
 \label{SZ1panel}
\end{figure}

\begin{figure}[t]
\centering{
\includegraphics[width=0.8\linewidth]{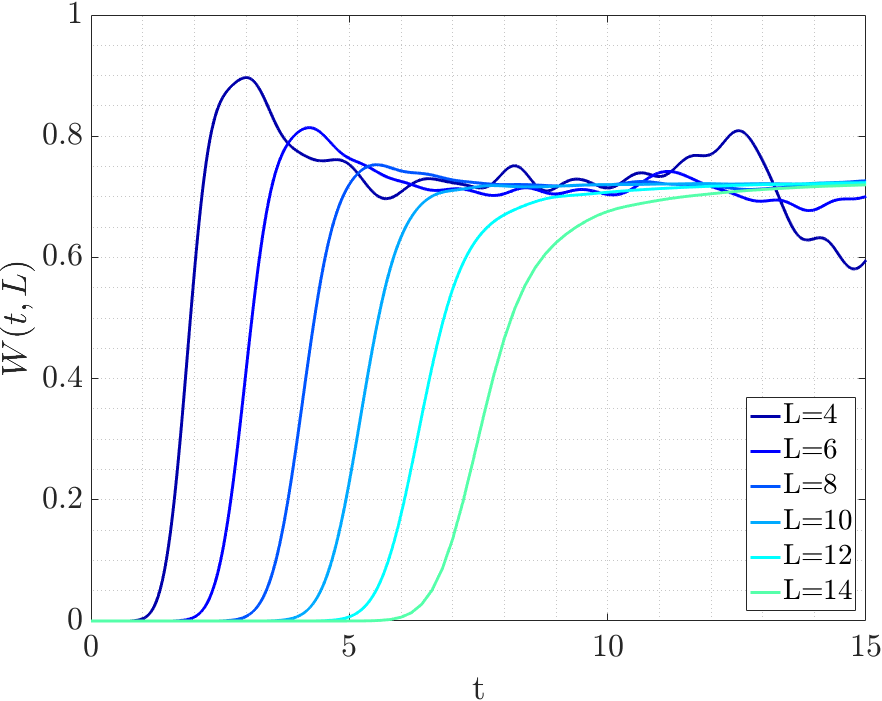}
}
\caption{
Pauli weight $W(L,t)$ for the operator $Z_1(t)$ at site $L$ as a function of time, for system sizes $L=4,6, \ldots, 14$.
}
 \label{fig:pauliweight}
\end{figure}

\begin{figure}[t]
\centering{
\includegraphics[width=0.8\linewidth]{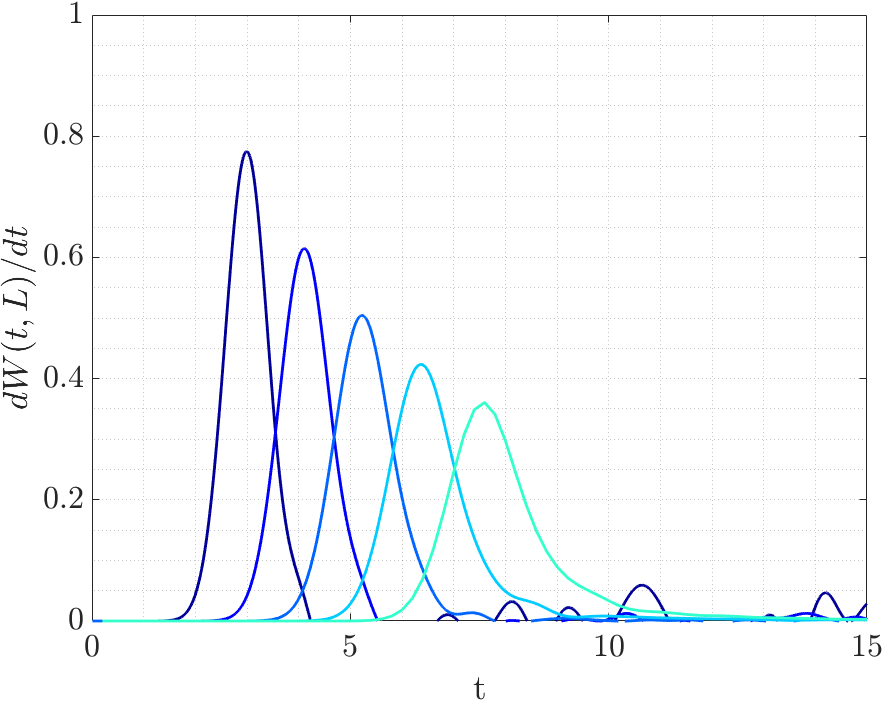}
}
\caption{
Derivative $\partial W(L,t)/\partial t$ of the Pauli weight in Fig.~\ref{fig:pauliweight} for system sizes $L=4,6, \ldots, 14$. Peaks are used to define a time $t_\text{arrival}$ at which $Z_1(t)$ spreads to the right hand end of the chain.
}
 \label{fig:pauliweightderivative}
\end{figure}

\begin{figure}[b]
\centering{
\includegraphics[width=0.8\linewidth]{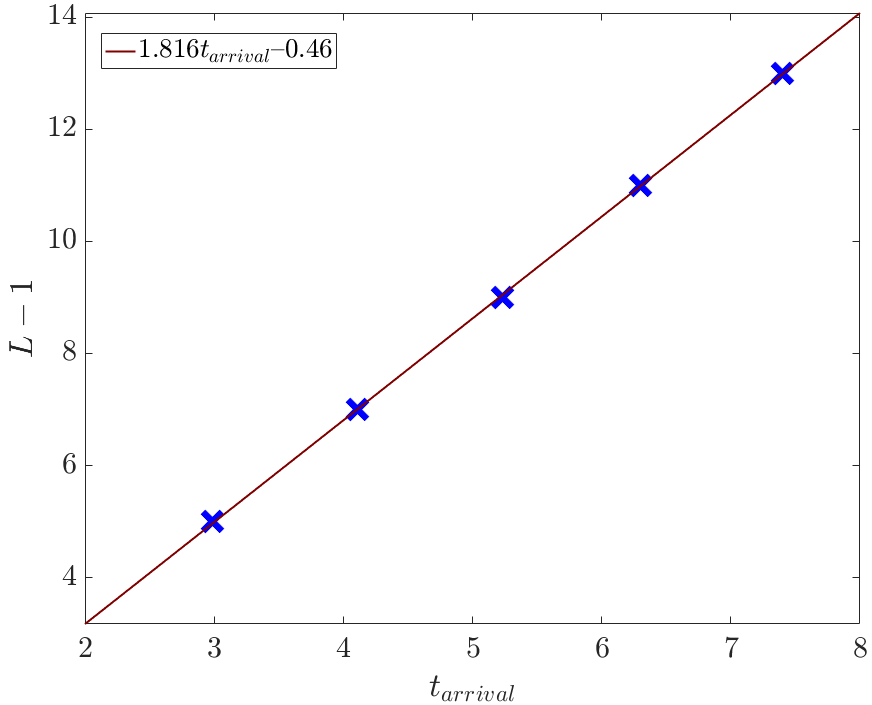}
}
\caption{
Estimate of $v_B$.
Plot shows distance $L-1$ between left and rightmost sites versus the time $t_\text{arrival}$ at which the operator $Z_1(t)$ reaches the rightmost site. This time is defined using the Pauli weight as in the text. The fit gives $v_B \simeq 1.82$.
}
 \label{fig:vBestimate}
\end{figure}

\section{Estimates of $\sspr$ and $v_B$}
\label{app:extrapolationsspr}

In this section we estimate the constants $\sspr$ and $v_B$ using the time evolution of the operator $Z_1(t)$. 

In Fig.~\ref{SZ1panel} we show the operator entanglement of $Z_1(t)$ in systems of size $L=8,\, 10, \, 12,\,14$. For each value of $L$ we obtain an estimate of $\sspr$ from the data at time $t=L/2$ by fitting the slope of the section of the curve shown in red. (We use the 3 or 4 $x$-values closest to the midpoint between the maximum of the curve and the last point.) Such estimates should converge to $\sspr$ as $L\rightarrow \infty$. We obtain
\be
0.74, \,\,\, 0.81, \,\,\, 0.85, \,\,\, 0.87 
\ee
for $L=8,\,10,\,12,\,14$ respectively. It is hard to estimate an error bar with this number of points, but extrapolating to ${L=\infty}$ suggests that $\sspr$ lies in the range \protect{$0.9$---$1.0$}.

To estimate $v_B$ we use the ``Pauli weight'' $W(L,t)$, which is the weighted fraction of the Pauli strings contributing to $Z_1(t)$ that have support at site $L$ \cite{rss}. This quantity becomes appreciable when the front of the operator reaches the end of the chain at site $L$: see Fig.~\ref{fig:pauliweight}.
For each $L$, we define the arrival time $t_\text{arrival}$ of the front as the time at which $\partial W(L,t)/\partial t$, plotted in Fig.~\ref{fig:pauliweightderivative},  is maximal. We can extract $v_B$ from the relation  ${L \simeq v_B t_\text{arrival} + c}$ which should hold at late times ($c$ is a constant). In principle, the advantage of this protocol is that finite size effects are controlled only by the size of $L$ (if instead we fix $L$ and study the front at various positions $x$, finite size effects are controlled by the size of $L$, $x$ and $L-x$).

In Fig.~\ref{fig:vBestimate} we plot $t_\text{arrival}$ against the system size.  Fitting the data to ${L = v_B t_\text{arrival} + c}$ gives $v_B\simeq 1.82$. 
The uncertainty in this estimate is larger than the quality of the fit suggests: the significant broadening of the front visible in Figs.~\ref{fig:pauliweight},~\ref{fig:pauliweightderivative} means that the estimate of $v_B$ will be sensitive to the way in which $t_\text{arrival}$ is defined (see \cite{ms} for a related discussion).

%

\end{document}